\RequirePackage{ifpdf}
\ifpdf 
\documentclass[pdftex]{sigma}
\else
\documentclass{sigma}
\fi

\numberwithin{equation}{section}

\begin{document}

\allowdisplaybreaks

\renewcommand{\thefootnote}{$\star$}

\renewcommand{\PaperNumber}{064}

\FirstPageHeading

\ShortArticleName{Hidden Symmetry from Supersymmetry}

\ArticleName{Hidden Symmetry from
Supersymmetry\\ in One-Dimensional Quantum Mechanics\footnote{This paper is a contribution to the Proceedings of the VIIth Workshop ``Quantum Physics with Non-Hermitian Operators''
   (June 29 -- July 11, 2008, Benasque, Spain). The full collection
is available at
\href{http://www.emis.de/journals/SIGMA/PHHQP2008.html}{http://www.emis.de/journals/SIGMA/PHHQP2008.html}}}

\Author{Alexander A. ANDRIANOV~$^{\dag\ddag}$ and Andrey V. SOKOLOV~$^\dag$}

\AuthorNameForHeading{A.A. Andrianov and A.V. Sokolov}

\Address{$^\dag$~V.A.~Fock Department of Theoretical Physics,
Sankt-Petersburg State University,\\
\hphantom{$^\dag$}~198504 St. Petersburg, Russia}
\EmailD{\href{mailto:andrianov@bo.infn.it}{andrianov@bo.infn.it}, \href{mailto:avs_avs@rambler.ru}{avs\_avs@rambler.ru}}

\Address{$^\ddag$~Departament ECM and  ICCUB, Departament de Fisica, Universitat de Barcelona,\\
\hphantom{$^\ddag$}~08028 Barcelona, Spain}

\ArticleDates{Received March 05, 2009, in f\/inal form June 02, 2009;  Published online June 17, 2009}

\Abstract{When several inequivalent supercharges form a closed superalgebra in Quantum Mechanics it entails the appearance of hidden symmetries of a Super-Hamiltonian. We exa\-mi\-ne this problem in one-dimensional QM for the case of periodic potentials and potentials with f\/inite number of bound states. After the survey of the results existing in the subject
the algebraic and analytic properties of hidden-symmetry dif\/ferential operators are rigorously elaborated in the Theorems and illuminated by several examples.}


\Classification{81Q60; 81R15; 47B15}

\renewcommand{\thefootnote}{\arabic{footnote}}
\setcounter{footnote}{0}

\section{Introduction}

The interest to hidden  symmetries in quantum dynamical systems accompanies
Quantum Mechanics (QM) nearly since its very formulation~\cite{fock}. If
existing they play an exceptional role  in unraveling the energy spectra and,
especially, in construction of integrable systems related to the soliton
dynamics~\cite{teorsol} where hidden symmetry operators represent key ingredients of the Lax pair method. One should also mention the related study of conditional (formal) symmetries of
the Shr\"odinger equation realized by dif\/ferential operators of f\/inite order which was undertaken in~\cite{nikitin}.

Recently a particular class of hidden symmetries in one-dimensional
QM has been explored with the help of Non-linear Supersymmetric QM
\cite{ansok}. The idea that a hidden QM symmetry  is accounted for
by the existence of several supercharges was outlined  in
\cite{acin1}. Accordingly  such a~connection leads to the
realization of ${\cal N}=4$ Non-linear SUSY QM with a central
charge. In~\cite{ansok} the representation of this algebra was found
among the quantum systems with ref\/lectionless potentials.

A similar class of SUSY-induced hidden symmetries exists among periodic  systems, its
possibility for periodic potentials was guessed in~\cite{kiev}. In
\cite{braden,dunnefei} it was found that a conventional supersymmetric extension of a
periodic quantum system may give an  isospectral pair with a~zero energy doublet of the
ground states. For non-periodic systems with real scalar potentials it can happen
only\footnote{ There are more options for matrix \cite{acin2} and complex potentials \cite{andcansok07}.}
for non-linear SUSY with supercharges of higher order in derivatives~\cite{acdi}. It
was also established~\cite{dunnefei} that among isospectral supersymmetric periodic
systems there are some self-isospectral samples for which the partner potentials are
identical in shape but related by translation for a half-period and/or  ref\/lection.
Later on, dif\/ferent aspects of isospectral and self-isospectral supersymmetric periodic
systems were examined in~\cite{khasump,ferneni,conipl}. Recently~\cite{djferc} a~property of the self-isospectrality was revealed in some periodic f\/inite-gap systems
based on a nonlinear supersymmetry of the second order. In the papers~\cite{cojapl08},
the superextension of quantum periodic systems has been studied with a parity-even
f\/inite-gap potential of general form, and it was shown that it is characterized by a
tri-supersymmetric structure. This supersymmetric structure originates from the higher
order dif\/ferential operator of the Lax pair.

\looseness=1
After this concise review of the state-of-art let us formulate the aim of our paper and
its novel results. The main goal is to f\/ill the missing points of previous studies of
hidden symmetry operators as elements of  a SUSY algebra, in other words, when they
interplay with the Darboux--Crum construction. We pay more attention to the spectral
problem and therefore investigate the kernels of both intertwining and symmetry
operators simultaneously. The novel element is in extension of present and previously
known facts to complex potentials~\cite{andcansok07, ancandi,mostafa,samroy,quasi,bender,levai,znojil,dorey} for Hamiltonians which
are not necessarily diagonalizable~\cite{andcansok07,mostafa}. We examine also the
possibility of a non-Hermitian Hamiltonian to be $PT$ symmetric~\cite{bender}. Section~\ref{section2} is devoted to the summary of algebraic and dif\/ferential properties of Non-linear SUSY
in one-dimensional QM (in particular, of the matrix $\bf S$ construction) and to the
def\/initions related to minimization of this algebra while keeping invariant the
spectrum and the structure of the Super-Hamiltonian.  In Section~\ref{section3} Extended SUSY is
introduced and its optimal structure is elucidated. When the extension is nontrivial
its central charge is built of  symmetry operators which can be chosen to be
antisymmetric under transposition ($t$-antisymmetric) by a suitable redef\/inition of
elements of supercharges. When the Hamiltonians are $PT$-symmetric the corresponding
$t$-antisymmetric symmetry operators become $PT$-antisymmetric. Theorem~\ref{theorem3} culminating
for the paper is formulated and proven in Section~\ref{section4}. It summarizes all about
$t$-antisymmetric non-minimizable symmetry operators for periodic potentials, including
the smoothness of their coef\/f\/icients, the content of their kernels based on the
characteristic polynomials, relationship of zero-modes and the borders of forbidden
bands, their factorization into the product of elementary intertwining operators and
generation of the ladder of intermediate Hamiltonians together with corresponding
symmetry operators intertwined with the basic ones. Separately, in Section~\ref{section5} we re-analyze
the Hamiltonians with bound states and extend the previous results to complex
potentials for Hamiltonians which may possess not only eigenfunctions but also
normalizable associated functions. Conventionally they are not diagonalizable but can
be reduced to a Jordan form. In this section, mainly the potentials with constant
asymptotics are within the scope although the basic results are valid for a large class
of potentials.  Again the content of the symmetry operator kernels based on the
characteristic polynomials, relationship of their zero-modes and bound-states, their
factorization into the product of elementary intertwining operators and generation of
the ladder of intermediate Hamiltonians together with their symmetry operators are
investigated. A number of examples illuminating the general statements are elaborated
throughout the paper. In Conclusions possible generalizations of the obtained results
onto the Hamiltonians with quasiperiodic potentials and the Hamiltonians with
ref\/lectionless potentials against the background of f\/inite-zone potentials are brief\/ly
discussed.

\section{Basic def\/initions and notations}\label{section2}

We start with the  def\/inition of a  SUSY algebra in Quantum Mechanics \cite{nico,witten,coop1,abi,nieto,miel,fern,abei, sukum,genden} and notations for
its components. Consider two one-dimensional Hamiltonians of the
Schr\"odinger type def\/ined on the entire axis
and having suf\/f\/iciently smooth and in general complex-valued
potentials $V_1(x)$ and $V_2(x)$. The Hamiltonians are assembled into
the Super-Hamiltonian, \[ H =
\begin{pmatrix}h^+& 0\\ 0& h^-\end{pmatrix},\qquad h^+=-\partial^2+V_1(x),\quad h^-=-\partial^2+
V_2(x),\qquad\partial\equiv d/dx .\]
Assume that the Hamiltonians $h^+$ and $h^-$ have  (almost) equal energy spectra of
bound states and equal spectral densities of the continuous spectrum part.  Let it be
provided by the Darboux--Crum operators $q^\pm_N$ with the help of intertwining:
\begin{gather}
h^+q^+_N = q^+_Nh^-,\qquad q^-_Nh^+ = h^-q^-_N.\label{int2}
\end{gather} Further on, we restrict
ourselves to dif\/ferential Darboux--Crum operators of a f\/inite order~$N$,
\[ q^\pm_N =
\sum^N_{k=0} w^\pm_k (x)\partial^k,\qquad w^\pm_N\equiv(\mp1)^N, \] with suf\/f\/iciently
smooth and in general complex-valued coef\/f\/icients $w^\pm_k (x)$. In this case, in the
fermion number representation, the nonlinear ${\cal N} = 1$ SUSY QM
\cite{acdi,bagsam97,ancan,ast,ast2} is formed by
means of the nilpotent su\-per\-char\-ges: \[ Q_N = \begin{pmatrix}0& q^+_N\\
0& 0\end{pmatrix},\qquad \bar Q_N = \begin{pmatrix} 0& 0\\ q^-_N&
0\end{pmatrix},\qquad Q^2_N = \bar Q^2_N = 0. \] Obviously, the
intertwining relations \eqref{int2} lead to the supersymmetry of the
Hamiltonian~$H$: \[[H,Q_N] = [H,\bar Q_N] = 0.\]
In view of intertwining \eqref{int2} the kernel of $q_N^\pm$ is an
invariant subspace with respect to the Hamiltonian $h^\mp$,
\[ h^\mp\ker q_N^\pm\subset \ker q_N^\pm.\]
Hence, there is
a constant $N\times N$ matrix ${\bf S^\mp}\equiv\| S^\mp_{ij}\|$ for an
arbitrary basis $\phi^\pm_1(x)$, \dots, $\phi^\pm_N(x)$ in $\ker
q_N^\pm$ such that
\begin{gather}
h^\mp\phi^\pm_i=\sum_{j=1}^NS^\mp_{ij}\phi^\pm_j.\label{h9}\end{gather} In what
follows, for an intertwining operator, its {\it matrix $\bf S$} is
def\/ined as the matrix which is related to the operator in the same
way as the matrices $\bf S^\mp$ are related to $q^\pm_N$. In this
case, we do not specify the basis in the kernel of the intertwining
operator in which the matrix $\bf S$ is chosen if we concern ourselves only with
spectral characteristics of the matrix, or, what is the same,
spectral characteristics of the restriction of the corresponding
Hamiltonian to the kernel of the intertwining operator considered
(cf.~\eqref{h9}).

The introduced above, nonlinear SUSY algebra  is closed by the following
relation between the supercharges and Super-Hamiltonian:
\[\{Q_N,\bar Q_N\} ={\cal P}_N(H),\] where ${\cal P}_N(H)$ is a
dif\/ferential operator of $2N$th order commuting with the
Super-Ha\-mil\-to\-ni\-an. Depending on a relation between the
supercharges $Q_N$ and $\bar Q_N$ (the intertwining operators
$q^\pm_N$), the operator $P_N(H)$ can be either a polynomial of the
Super-Hamiltonian, if the supercharges are connected by the
operation of transposition: \begin{gather} \bar Q_N = Q^t_N, \qquad
q_N^-=(q_N^+)^t\equiv\sum_{k=0}^N(-\partial)^kw_k^+(x),\label{trans}\end{gather}
or in general a function of both the Super-Hamiltonian and a
dif\/ferential symmetry operator of odd order in derivatives (see a
detailed analysis and references in~\cite{ansok}). In the present paper, we
conf\/ine ourselves to the f\/irst case in which the conjugated
supercharge is produced by transposition \eqref{trans}. A relevant
theorem on the structure of such a SUSY \cite{ansok,andcansok07}
reads.

\begin{theorem}[on SUSY algebra with transposition
symmetry]\label{theorem1} The closure of the supersymmetry algebra with $\bar
Q_N=Q_N^t$ takes a polynomial form: \[ \{Q_N,Q^t_N\} = \det[E{\bf
I}-{\bf S}^+]_{E=H} = \det[E{\bf I}-{\bf S}^-]_{E=H}\equiv {\cal
P}_N(H),\] where $\bf I$ is the identity matrix and $\bf S^\pm$ is
the
matrix $\bf S$ of the intertwining operator $q_N^\mp$.
\end{theorem}

\begin{corollary} The spectra of the matrices $\bf S^+$ and $\bf
S^-$ are equal.
\end{corollary}

A basis in the kernel of an intertwining operator in which the
matrix $\bf S$ of this operator has a Jordan form is called {\it
canonical}; elements of a canonical basis are called {\it
trans\-for\-ma\-ti\-on functions}.

If a Jordan form of the matrix $\bf S$ of an intertwining operator
has cells of a size higher than one, then the corresponding canonical
basis contains not only formal solutions of the Schr\"odinger
equation but also formal associated functions, which are def\/ined as
follows.

A function $\psi_{n,i}(x)$ is called {\it a formal associated
function of $i$-th order} of the Hamiltonian $h$ for a spectral
value $\lambda_n$ if
\[(h-\lambda_n)^{i+1}\psi_{n,i}\equiv0\qquad
\text{and} \qquad (h-\lambda_n)^i\psi_{n,i}\not\equiv0.\] The term
``formal'' emphasizes that this function is not necessarily
normalizable (not necessa\-ri\-ly belongs to $L_2({\mathbb R}))$. In
particular, an associated function $\psi_{n,0}(x)$ of zero order is
a formal eigenfunction of $h$ (not necessarily a normalizable
solution of the homogeneous Schr\"odinger equation).

Assume that the intertwining operator $q^\pm_N$ is represented as a
product of intertwining operators $k^\pm_{N-M}$ and $p^\pm_M$, $0 <M
<N$, so that \begin{gather*} q^+_N = p^+_Mk^+_{N-M},\qquad q_N^- =
k^-_{N-M}p^-_M,\qquad p^+_Mh_M = h^+p^+_M,\qquad p^-_Mh^+ =
h_Mp^-_M,\nonumber\\  k^+_{N-M}h^- = h_Mk^+_{N-M},\qquad k^-_{N-M}h_M =
h^-k^-_{N-M},\qquad \text{and} \qquad h_M =-\partial^2 + v_M(x),\end{gather*}
where the coef\/f\/icients $k^\pm_{N-M}$ and $p^\pm_M$ as well as the
potential $v_M(x)$ may be complex-valued and/or containing  pole-like singularities. The
Hamiltonian $h_M$ is called {\it intermediate with respect to $h^+$
and $h^-$}. In this case, by Theorem~\ref{theorem1}, the spectrum of the matrix
$\bf S$ of the operator $q^\pm_N$ is the union of the spectra of the
matrices $\bf S$ for the operators $k^\pm_{N-M}$ and $p^\pm_M$.

The potentials $V_1(x)$ and $V_2(x)$ of the Hamiltonians $h^+$ and
$h^-$ are interrelated by the equation \begin{gather} V_2(x) = V_1(x)-2[\ln
W(x)]'',\label{v2v1}\end{gather} where $W(x)$ is the Wronskian of elements of an
arbitrary (a canonical as well) basis in $\ker q^-_N$. The validity
of equation~\eqref{v2v1} follows from the Liouville--Ostrogradsky relation
and the equality of coef\/f\/icients at $\partial^N$ in $q^-_Nh^+$ and
$h^-q^-_N$ (see the intertwining in~\eqref{int2}).

With the help of a basis $\phi^\pm_1(x)$, \dots, $\phi^\pm_N(x)$ in
$\ker q_N^\pm$ the operator $q_N^\pm$ can be presented in the form
\begin{gather}
q_N^\pm=\frac{1}{W_\pm(x)}\begin{vmatrix}\phi_1^\pm(x)&\phi_1^{\pm\prime}(x)&\dots&
\phi_1^{\pm(N)}(x)\\
\hdotsfor{4}\\
\phi_N^\pm(x)&\phi_N^{\pm\prime}(x)&\dots&\phi_N^{\pm(N)}(x)\\
1&\partial&\dots&\partial^N\end{vmatrix},\label{qpres}\end{gather} where
$W_\pm(x)$ is the Wronskian of these basis elements and the dif\/ferential operators
must be placed on the right-hand side when calculating the determinant elements.

An intertwining operator $q^\pm_N$ is called {\it minimizable} ($q_N^\pm$ {\it
can be ``stripped off''}) if this operator can be represented in the
form\footnote{The possibility of existence of a
cofactor polynomial in the Hamiltonian for an intertwining operator was mentioned  in
\cite{bagsam97}.} \[ q^\pm_N = {\cal P}(h^\pm)p^\pm_M = p^\pm_M{\cal
P}(h^\mp),\]
where $p^\pm_M$ is an operator of order $M$ which intertwines
the same Hamiltonians as $q^\pm_N$ (i.e., $p^\pm_Mh^\mp= h^\pm
p^\pm_M$), and ${\cal P}(h)$ is a polynomial of degree $(N-M)/2>0$.
Otherwise, the intertwining operator $q^\pm_N$ is called {\it non-minimizable}
($q_N^\pm$ {\it cannot be ``stripped off''}). The following theorem~\cite{ansok,andcansok07} contains the necessary and suf\/f\/icient conditions controlling whether
an intertwining
operator is minimizable or not.

\begin{theorem}[on minimization of an intertwining operator]\label{theorem2}
An intertwining ope\-ra\-tor~$q^\pm_N$ can be represented in the form \begin{gather*}
q^\pm_N = p^\pm_M\prod^m_{l=1} (\lambda_l-h^\mp)^{\delta k_l},\end{gather*}
where
$p^\pm_M$ is a non-minimizable operator intertwining the same Hamiltonians as
$q^\pm_N$ $($so that $p^\pm_M h^\mp=h^\pm p^\pm_M)$, if and only if a Jordan
form of the matrix $\bf S$ of the operator $q^\pm_N$ has $m$ pairs $($and no more$)$
of Jordan cells with equal eigenvalues $\lambda_l$ such that, for the
$l$-th pair, $\delta k_l$ is an order of the smallest cell and $k_l +\delta
k_l$ is an order of the largest cell. In this case, $M = N-2\sum^m_{l=1}
\delta k_l =\sum^n_{l=1} k_l$ $($where the $k_l$, $m + 1\leqslant
l\leqslant n$, are orders of the remaining unpaired Jordan cells$)$.
\end{theorem}

\begin{remark} A Jordan form of the matrix $\bf S$ of the
intertwining operator $q^\pm_N$ cannot have more than two cells with
the same eigenvalue $\lambda$; otherwise, $\ker (\lambda-h^\mp)$
includes more than two linearly independent elements.\end{remark}

\begin{corollary} Jordan forms of the matrices $\bf S$ of the operators
$q^+_N$ and $q^-_N$ coincide up to permutation of Jordan cells.
\end{corollary}

\section{Several supercharges and extended SUSY}\label{section3}

Let us examine the case when for a Super-Hamiltonian $H$ there are two
dif\/ferent supercharges~$K$ and $P$ of the type $Q$ and of the order $N$ and
$M$ respectively, \[ K = \left(\begin{array}{cc}
0 &  k^+_{N}\\
0 &  0
\end{array}\right),\qquad P
= \left(\begin{array}{cc}
0 & p^+_{M} \\
0 & 0
\end{array}\right).
\] In particular, if a  complex supercharge $Q$ exists for the Hermitian
Super-Ha\-mil\-to\-ni\-an $H$, we can choose $K$ and $P$ as $(Q+Q^*)/2$ and
$(Q-Q^*)/(2i)$ respectively, where $^*$ denotes complex conjugation of
coef\/f\/icient functions. Let us assume that $N>M$ (in the case $N=M$, instead of~$p_{M}^+$ we can use
 a suitable linear combination of $k_N^+$ and $p_{M}^+$ ,
the order of which is less than~$N$). Each of supercharges $K$ and $P$
generates a unique supercharge of the type $\bar Q$:  \[ \bar K = K^t =
 \begin{pmatrix} 0 &  0 \\   k^-_{N}  &  0
\end{pmatrix}, \qquad  k_N^- = (k_N^+)^t; \qquad  \bar
P = P^t =   \begin{pmatrix} 0 &  0 \\  p^-_{M}  &
 0
\end{pmatrix}, \qquad  p_{M}^- = (p_{M}^+)^t .  \]

The existence of two supercharges of the type $Q$ (i.e.\ $K$ and $P$)
conventionally signif\/ies the extension of SUSY algebra. To close the algebra
one has to include all anti-commutators between supercharges. Two supercharges
$K$ and~$P$ generate two Polynomial SUSY, \begin{gather*} \left\{K, K^t\right\} =
\tilde{\cal P}_N (H),\qquad \left\{P, P^t \right\} = \tilde{\cal P}_{M} (H),
\end{gather*} which have to be embedded into a ${\cal N} =4$ SUSY algebra. The
closure of the extended, ${\cal N} =4$ SUSY algebra is given by
\[
\left\{P, K^t \right\} \equiv {\cal R} =  \begin{pmatrix}
 p^+_{M} k^-_{N} &  0\\
0 & k^-_{N} p^+_{M}
\end{pmatrix},\qquad
\left\{K, P^t\right\} \equiv \bar{\cal R} =  \begin{pmatrix}
k^+_{N} p^-_{M} &  0\\
0 & p^-_{M} k^+_{N}
\end{pmatrix} .\]
Evidently the components of operators ${\cal R}$, $\bar{\cal R}=
{\cal R}^t$ are dif\/ferential operators of $N + M$ order commuting
with the Hamiltonians $h^\pm$, hence they form symmetry operators
${\cal R}$, $\bar{\cal R}$ for the Super-Hamiltonian. However, in
general, they are not polynomials of the Hamiltonians $h^\pm$ and
these symmetries impose certain constraints on potentials.

Let us f\/ind the formal relation between the symmetry operators
${\cal R}$, $\bar {\cal R}$ and the Super-Hamiltonian. These
operators can be decomposed into $t$-symmetric and $t$-antisymmetric
parts, \begin{gather*} {\cal B}\equiv \tfrac12({\cal R} +\bar {\cal R}) \equiv
 \begin{pmatrix}
b^+ & 0\\
0 & b^-
\end{pmatrix},\qquad
{\cal E} \equiv \tfrac12 ({\cal R} - \bar {\cal R})\equiv
 \begin{pmatrix}
e^+ & 0\\
0 & e^-
\end{pmatrix}. 
\end{gather*} The operator ${\cal B}$ plays essential role in the
one-parameter non-uniqueness of the SUSY algebra. Indeed, one can
always redef\/ine  the higher-order supercharge as follows, \begin{gather}
K_{\zeta} =  K + \zeta P,\qquad \left\{ K_{\zeta},
K^{t}_\zeta\right\} = \tilde{\cal P}_{\zeta,N} (H) \label{redef} \end{gather}
keeping the same order $N$ of Polynomial SUSY for arbitrary complex
parameter $\zeta$. From \eqref{redef} one gets, \[ 2 \zeta {\cal B}
(H) = \tilde{\cal P}_{\zeta,N} (H) -\tilde{\cal P}_N (H) - \zeta^2
\tilde{\cal P}_{M} (H), \] thereby $t$-symmetric operator ${\cal
B}$ is a polynomial of the Super-Hamiltonian of the order $N_b
\leqslant(N+M)/2$.

If the second $t$-antisymmetric symmetry operator~$\cal E$ does not vanish
identically, then it is a~dif\/ferential operator of {\it odd} order and cannot
be realized by a polynomial in~$H$. But at the same time
\begin{gather}
{\cal E}^2 (H)
 \equiv  \tfrac14\left[ ({\cal R} + \bar{\cal R})^2-2({\cal R}\bar{\cal
R} + \bar{\cal R} {\cal R}) \right] =  {\cal B}^2 (H) -  \tilde{\cal
P}_N (H) \tilde{\cal P}_{M} (H) \equiv  -{\cal P}_e(H) \label{secsym}
\end{gather} is a
polynomial in $H$. Thus the nontrivial operator ${\cal E} (H)$ is a non-polynomial
function of~$H$~-- the square root of \eqref{secsym} in an operator sense. This
operator is certainly non-trivial if the sum of orders $N + M$ of the
operators $k^\pm_N$ and $p^\pm_{M}$ is odd and therefore the order of ${\cal E} (H)$ amounts to $N_e = N + M$.

As it is known~\cite{ansok,andcansok07} the case ${\cal E}(H)=0$ implies
relationship between the supercharges $K$ and $P$, namely, their identity
after minimization. The case ${\cal B}(H)=0$ entails the equalities \[ {\cal
E}= {\cal R} \ \Leftrightarrow \ {\cal R}=-\bar{\cal R},\] accordingly the
Hamiltonian $h^-$ ($h^+$) is intermediate with respect to some
fac\-to\-ri\-za\-ti\-on of the nontrivial $t$-antisymmetric symmetry operator
$e^+$ ($e^-$) of the Hamiltonian $h^+$ ($h^-$), \[
e^+=p_M^+k_N^-(=-k_N^+p_M^-)\qquad \big(e^-=k_N^-p_M^+(=-p_M^-k_N^+)\big).\]
If the supercharges $K$ and $P$ are independent (i.e.\ if ${\cal
E}(H)\ne0$) one can achieve vanishing of ${\cal B}(H)$-type symmetry operator
taking instead of $K$ and $P$ the new pair of independent supercharges: \[
\tilde K={\cal P}_M(H)K-{\cal
B}(H)P=\begin{pmatrix}0&{\cal P}_M(h^+)k_N^+-b^+p_M^+\\0&0\end{pmatrix},\qquad P,\]
wherefrom it follows that \begin{gather*}
\tilde{\cal B}(H)=\tfrac12 \big(\{P,\tilde
K^t\}+\{\tilde K,P^t\}\big) \\
\phantom{\tilde{\cal B}(H)}{} =
\tfrac12\big({\cal P}_M(H)\{P,K^t\}-{\cal B}(H)\{P,P^t\}+{\cal
P}_M(H)\{K,P^t\}-{\cal B}(H)\{P,P^t\}\big)
\\
\phantom{\tilde{\cal B}(H)}{}
= {\cal P}_M(H){\cal B}(H)-{\cal B}(H){\cal P}_M(H)=0.
\end{gather*} If the supercharges $K$ and $P$ are
dependent (i.e.\  if ${\cal E}(H)=0$) then obviously the
po\-ly\-no\-mial~${\cal B}(H)$ cannot vanish and its order is $(N+M)/2$.

Let us assume that ${\cal E}(H)\ne0$ and  minimize the symmetry
operators $e^+$ and $e^-$, \[ e^+={\cal P}_+(h^+)\tilde e^+,\qquad e^-={\cal
P}_-(h^-)\tilde e^-,\] where $\tilde e ^+$ and $\tilde e^-$ are
non-minimizable symmetry operators for $h^+$ and $h^-$ respectively. The
operators $\tilde e^+$ and $\tilde e^-$ are $t$-antisymmetric as well, because
in the opposite case \[ e^\pm +(e^\pm)^t= {\cal P}_\pm(h^\pm)\tilde e^\pm
+(\tilde e^\pm)^t{\cal P}_\pm(h^\pm)={\cal P}_\pm(h^\pm)[\tilde e^\pm+(\tilde
e^\pm)^t]\ne0.\] It is known (see the example in \cite{ansok}) that the
polynomials ${\cal P}_+$ and ${\cal P}_-$ are  in general dif\/ferent. Thus
evidently the symmetry operator ${\cal E}(H)$ of the SUSY algebra can be minimized only by
separation of the polynomial in the Super-Hamiltonian $H$ which is the
greatest common polynomial divisor of the polynomials ${\cal P}_+$ and ${\cal
P}_-$. In view of
Theorem~\ref{theorem1} and $t$-antisymmetry of $\tilde{\cal E}(H)$, the spectra of the matrices
$\bf S$ for elements of the
minimized ${\cal E}(H)$ (we shall denote it by $\tilde{\cal E}(H)$)  are identical among
themselves and to the set of zeros of the polynomial $\tilde{\cal E}^2(H)$.
Any element of these spectra obviously belong either to the spectrum of the
matrix $\bf S$ of $\tilde e^+$ or to the spectrum of the matrix~$\bf S$ of~$\tilde e^-$.

In the following sections it will be shown  that the spectrum of the
matrix $\bf S$ of a non-minimizable $t$-antisymmetric  operator $e$ for a
Hamiltonian $h$ consists of energies of all bound states of $h$ and of all
boundaries of continuous spectrum of $h$ as well as (in the case of
non-Hermitian $h$) of other characteristic points of the $h$ spectrum
(herein under $h$ and $e$ we imply any of Hamiltonians $h^\pm$ and of a related, properly minimized symmetry operator $\bar e^\pm$). Hence, all
zeros of the polynomial $\tilde{\cal E}^2(H)$ possess a physical meaning and
represent characteristic points of $H$ spectrum, in particular, all energies
of bound states, all boundaries of continuous spectrum etc.

It has been established \cite{ansok,andcansok07} that any nonzero $t$-antisymmetric
symmetry operator can be presented in the form $ {\cal P}(h)e,$ where
$\cal P$ is a polynomial of the Hamiltonian and $e$ is
a unique $t$-antisymmetric non-minimizable symmetry operator with unit
coef\/f\/icient at the highest-order derivative. Moreover, if the potential
$V(x)$ is real-valued then all coef\/f\/ici\-ents of $e$ are obviously real-valued
as well. The two following sections are devoted to investigation of properties of the
operator $e$.

For the characteristic polynomial of the matrix $\bf S$ for $e$ (with
the help of which the $e$ squared is expressed  through the Hamiltonian $h$ in virtue of Theorem~\ref{theorem1}), we shall use the following notation \begin{gather} {\cal
P}_e(h)\equiv ee^t=-e^2=e^te.\label{harpol}\end{gather} It is evident that the
degree of this polynomial is equal to the order of $e$ and that $e$
is an algebraic function (square root of polynomial) of the Hamiltonian
$h$.

 We shall proceed in investigation of $e$ in two cases: in the case of
periodic $V(x)$ and in the case, when there are bound states for
$h$.

\begin{remark} One can easily check, that in the case of
$PT$-symmetric potential $V(x)$ the operator $PTePT$ is
$t$-antisymmetric dif\/ferential symmetry operator for $h$ of the same
order as $e$. Thus, in view of the uniqueness of $e$, its odd order
and of the equality $PT\partial=-\partial PT$ the following
relations hold: \begin{gather} PTePT=-e\ \Leftrightarrow\  PT e=-e
PT.\label{pte}\end{gather} It follows obviously from \eqref{pte} and from the
equality $PTh=hPT$ (which is equivalent to $PT$-symmetry of $V(x)$),
that:

(1) the part of a canonical basis in the $\ker e$, corresponding to
real eigenvalues of the matrix~$\bf S$ of~$e$, can be constructed
from $PT$-symmetric functions;

(2) if there is a non-real eigenvalue of the matrix $\bf S$ of $e$, then there
is also the complex conjugated eigenvalue of the same algebraic multiplicity
for this matrix and  the elements of a~canonical
basis in $\ker e$ corresponding to these eigenvalues can be constructed from mutually $PT$ conjugated functions.

As it is shown in the following sections, any eigenvalue
of the matrix $\bf S$ of $e$ is a characteristic point of $h$
spectrum. Thus in the case of unbroken $PT$-symmetry all elements of
a canonical basis in $\ker e$ can be chosen $PT$-symmetric.\end{remark}

\section[$t$-antisymmetric symmetry operators: Hamiltonians with periodic potential]{$\boldsymbol{t}$-antisymmetric symmetry operators:\\ Hamiltonians with periodic potential}\label{section4}

Properties of $t$-antisymmetric symmetry operator in
this case are elucidated in the following

\begin{theorem}\label{theorem3}Assume that:

$(1)$ the potential $V(x)$ of the Hamiltonian $h=-\partial^2+V(x)$
is a real-valued periodic function belonging to $C^\infty_{\mathbb R}$
and $X_0>0$ is a period of $V(x)$;

$(2)$ there is  a $t$-antisymmetric non-minimizable symmetry operator \[
e=\partial^N+\alpha_{N-1}(x)\partial^{N-1}+\cdots+\alpha_1(x)\partial+\alpha_0(x)
\] for the Hamiltonian $h$, \[ eh=he,\qquad
e^t
=-e,\] and $\alpha_l(x)$ belongs to $C^l_{\mathbb R}\cap C^2_{\mathbb
R}$, $l=0, \dots, N-1$;

$(3)$ $\psi_j(x)$ is a real-valued periodic or antiperiodic wave
function of $h$ corresponding to the boundary $E_j$ $(j+1$th from below$)$
between forbidden and allowed bands of the $h$ spectrum, $j=0,$ $1,
2, \dots$;

$(4)$ ${\cal P}_e(h)=e^te$.

\noindent Then:

$(1)$ $\alpha_l(x)$ is a real-valued periodic $($with the period $X_0)$
function belonging to $C^\infty_{\mathbb R}$, $l=0, \dots, N-2$ and
$\alpha_{N-1}(x)\equiv0$;

$(2)$ the following equalities hold, \begin{gather} e\psi_j=0,\qquad {\cal
P}_e(E_j)=0\Leftrightarrow {\cal P}_e(E_j)\psi_j={\cal
P}_e(h)\psi_j=-e^2\psi_j=0,\qquad j=0, 1, 2, \dots,  \label{p30}\end{gather} and
moreover:

$(a)$ the set of functions $\psi_j(x)$, $j=0, 1, 2$, \dots is a
canonical basis in $\ker e$;

$(b)$ any of the numbers $E_j$, $j=0, 1, 2, \dots$ is an
eigenvalue of algebraic multiplicity $1$ for the matrix $\bf S$ of
the operator $e$ and there are no other eigenvalues of this matrix;

$(3)$ there are $((N+1)/2)!$ $($and no more$)$ different nonsingular
factorizations of $e$ into product of one intertwining operator of the first
order and $(N-1)/2$ intertwining operators of the second order; moreover:

$(a)$ all intermediate Hamiltonians of these factorizations possess
the same spectrum as $h$ and potentials of all these Hamiltonians
are real-valued periodic $($with the period $X_0)$ functions belonging to
$C^\infty_{\mathbb R}$;


$(b)$ the coefficient at the highest-order derivative   in any intertwining
operator of the first or the second orders is $1$ and all other coefficients
of these operators are real-valued periodic $($with the period $X_0)$ functions
belonging to $C^\infty_{\mathbb R}$;

$(c)$ the spectrum of the matrix $\bf S$ of an intertwining operator of the
f\/irst order consists of $E_0$ and the spectrum of the matrix $\bf S$ of an
intertwining operator of the second order consists of borders of a
forbidden band 
so that every forbidden band corresponds to only one of the second order
operators;

$(d)$ if \begin{gather} e=r_{(N+1)/2}\cdots  r_1\label{e32}\end{gather} is one of the possible
factorizations of $e$ and $h_i$, $i=1, \dots, (N-1)/2$ are intermediate Hamiltonians corresponding to
this factorization, \[ r_{i}h_{i-1}=h_{i}r_i,\qquad
r_i^th_{i}=h_{i-1}r_i^t,\qquad i=1,\dots,(N+1)/2,\qquad h_0\equiv
h_{(N+1)/2}\equiv h,\] then a canonical basis in kernel of $r_i$ consists of
those band edge wave functions of $h_{i-1}$, energies of which form the
spectrum of the matrix $\bf S$ for $r_i$, and \begin{gather} r_i  \cdots
r_1\cdot r_{(N+1)/2}  \cdots   r_{i+1}\label{opsim}\end{gather} is a $t$-antisymmetric
non-minimizable symmetry operator of $N$-th
order for $h_i$, $i=1, \dots,$ $(N-1)/2$.
\end{theorem}

\begin{proof} Inclusion of the coef\/f\/icients of $e$ into $C^\infty_{\mathbb R}$ can
be proved on the same way as in Lemma~1 in~\cite{sok1}. Reality of these
coef\/f\/icients is obvious. The identity $\alpha_{N-1}(x)\equiv0$ holds in view
of $t$-antisymmetry of $e$. Periodicity of the coef\/f\/icients with the period
$X_0$ follows from the uniqueness of a normalized non-minimizable
$t$-antisymmetric symmetry operator and from the fact that operator dif\/ferent
from $e$ only by shift of all coef\/f\/icient's arguments by $X_0$ is a normalized
non-minimizable $t$-antisymmetric symmetry operator as well, by virtue of the
periodicity of the potential $V(x)$.

Let us now verify that the equalities~\eqref{p30} take place for any $j$. As all
coef\/f\/icients of $e$ are periodic and there is the only (up to a constant
cofactor) periodic or antiperiodic eigenfunction of $h$ for a border between
forbidden and allowed energy bands so $\psi_j(x)$ is eigenfunction of~$e$,
\[ e\psi_j=\mu_j\psi_j,\qquad j=0, 1, 2, \ldots,\] where $\mu_j$ is
corresponding eigenvalue. In view of periodicity of $e$ coef\/f\/icients,
$t$-antisymmetry of~$e$ and periodicity or anti-periodicity of $\psi_j(x)$ the
equalities holds,
\[ \mu_j\int_0^{X_0}\psi_j^2(x)\,dx=\int_0^{X_0}[e\psi_j](x)\psi_j(x)\,dx=
\int_0^{X_0}\psi_j(x)[e^t\psi_j](x)\,dx=
-\mu_j\int_0^{X_0}\psi_j^2(x)\,dx,\] wherefrom it follows in
view of reality of $\psi_j(x)$ that all numbers $\mu_j$, $j=0, 1,
2, \dots$ are equal to zero and thus the equalities \eqref{p30} are
valid for any $j$. It follows from \eqref{p30} for any $j$ and from
Theorem~\ref{theorem1} that all numbers $E_j$, $j=0, 1, 2, \dots$ belong to the
spectrum of the matrix $\bf S$ of $e$.

Let us show that the spectrum of the matrix $\bf S$ for $e$ contains the
values $E_j$, $j=0, 1, 2, \dots$ only. Suppose that the spectrum contains a
value $\lambda$, located either inside of an allowed band or inside of a
forbidden band or outside of real axis. Then this $\lambda$ in accordance with
Theorem~\ref{theorem1} is a zero of the polynomial ${\cal P}_e$. By virtue of periodicity
of $e$ coef\/f\/icients Bloch solutions of the equation $(h-\lambda)\psi=0$ are
formal eigenfunctions of the symmetry operator $e$ and moreover the
corresponding eigenvalues in view of the equalities $e^2=-{\cal P}_e(h)$ and
${\cal P}_e(\lambda)=0$ are zeros. Thus, the kernel of $e$ contains two
linearly independent solutions of the equation $(h-\lambda)\psi=0$ that
contradicts (see Theorem~\ref{theorem2}) to non-minimizability of $e$. Hence, the spectrum
of the matrix $\bf S$ of $e$ cannot contain a value situated inside of an
allowed or a forbidden band or outside of real axis.

Now suppose that the spectrum of the matrix $\bf S$ of $e$
contains a value~$\lambda$, located on a border between two
allowed bands. In this case~$\lambda$ is a zero of ${\cal P}_e$
again. In addition, in the case under consideration any two linearly
independent solutions of the equation $(h-\lambda)\psi=0$ are
simultaneously periodic or antiperiodic functions. It is evident
that acting of the operator $ie$ on elements of the kernel
$h-\lambda$ in some orthogonal basis with respect to scalar product
\[(f_1,f_2)=\int_0^{X_0}f_1(x)f_2^*(x)\,dx\] is described
by a Hermitian matrix. Consequently, a basis in the kernel of
$h-\lambda$ can be chosen from eigenfunctions of $e$. Together
with the condition ${\cal P}_e(\lambda)=0$ the latter
leads to
contradiction as before. Thus, the spectrum of the matrix
$\bf S$ of $e$ cannot contain a value situated on a border between
two allowed bands and this spectrum consists of the values
$E_j$, $j=0, 1, 2, \dots$ only.

Next we check that the algebraic multiplicity of any eigenvalue of the matrix
$\bf S$ of $e$ is one. After this check it will be obvious that the functions
$\psi_j(x)$, $j=0, 1, 2, \dots$ form a canonical basis in the kernel of $e$.
Suppose that the algebraic multiplicity of an eigenvalue $E_j$ of the matrix
$\bf S$ of $e$ is greater than one. It was shown in~\cite{dunnefei, khasump,
ferneni, djferc} that~$h$ can be intertwined with some Hamiltonian $\tilde h$, having real-valued periodic potential, with the help of intertwining
operator~$r$ of the f\/irst order (if $j=0$) or the second order (if $j>0$)
whose kernel consists of $\psi_0(x)$ (if $j=0$) or $\psi_j(x)$ and
$\psi_{j+1}(x)$ (if $j$ is odd) or $\psi_j(x)$ and $\psi_{j-1}(x)$ (if $j>0$
is even). Moreover, the Wronskian of transformation functions in any of these
cases has no zeros. Therefrom  as well as from~\eqref{v2v1}, \eqref{qpres} and the
condition $V(x)\in C^\infty_{\mathbb R}$ it follows that the potential of
$\tilde h$ and coef\/f\/icients of $r$ are real-valued and belong
to~$C^\infty_{\mathbb R}$.

\looseness=-1
Now consider the operator $rer^t$. It is obvious that this operator
is $t$-antisymmetric symmetry operator for $\tilde h$ and all its
coef\/f\/icients belong to $C^\infty_{\mathbb R}$. As $e$ is
$t$-antisymmetric  non-minimizable  symmetry operator, $\psi_j(x)$
belongs to a canonical bases in $\ker e$ and $\ker r$ and the
algebraic multiplicity of $E_j$ in the spectrum of the matrix $\bf
S$ of $e$ is greater than one, so that with the help of Lemma~1 from~\cite{ansok} one can separate from the right-hand side of $r$  the
intertwining operator $\partial-\psi'_j/\psi_j$ and from the
right-hand side of $e$ the same intertwining operator
$\partial-\psi'_j/\psi_j$ and simultaneously from the left-hand side
of $e$ the intertwining operator $(\partial-\psi'_j/\psi_j)^t$.
Thus, in view of Theorem~\ref{theorem1} the operator $rer^t$ is minimizable and
the polynomial which can be separated from $rer^t$ contains  the
binomial $E_j-\tilde h$ as a cofactor in the power which is greater
than or equal to two. Therefrom as well as from Theorems~\ref{theorem1} and~\ref{theorem2} and
from uniqueness of the normalized non-minimizable $t$-antisymmetric
symmetry operator $\tilde e$ for the Hamiltonian $\tilde h$ it
follows that algebraic multiplicity of~$E_j$ in the spectrum of the
matrix $\bf S$ of $\tilde e$ is less than the algebraic multiplicity
of $E_j$ in the spectrum of the matrix $\bf S$ of $e$, at least, by
two. Moreover, all coef\/f\/icients of $\tilde e$ are real-valued and
belong to $C^\infty_{\mathbb R}$, because in the opposite case
coef\/f\/icients of $rer^t$ obviously cannot be from $C^\infty_{\mathbb
R}$.

It can be verif\/ied that a canonical basis in $\ker r^t$ consists of periodic or
antiperiodic wave functions of the Hamiltonian $\tilde h$, namely: from $1/\psi_0(x)$
corresponding to the energy $E_0$ (if $j=0$) or from
$\psi_{j+1}(x)/[\psi'_j\psi_{j+1}-\psi_j\psi'_{j+1}]$ and
$\psi_j(x)/[\psi'_j\psi_{j+1}-\psi_j\psi'_{j+1}]$ corresponding to the energies $E_j$
and $E_{j+1}$ respectively (if $j$ is odd) or from
$\psi_{j-1}(x)/[\psi'_j\psi_{j-1}-\psi_j\psi'_{j-1}]$ and
$\psi_j(x)/[\psi'_j\psi_{j-1}-\psi_j\psi'_{j-1}]$ corresponding to the energies $E_j$
and $E_{j-1}$ respectively (if $j>0$ is even). Hence, the Hamiltonians $h$ and $\tilde
h$ as well as the symmetry operators $e$ and $\tilde e$ can be equally employed in the
previous argumentation. Thus, one f\/inds that if the algebraic multiplicity of~$E_j$ in
the spectrum of the matrix $\bf S$ of $\tilde e$ is greater than one, then the
algebraic multiplicity of $E_j$ in the spectrum of the matrix $\bf S$ of $e$ as
compared to itself is less, at least, by four. As well if the algebraic multiplicity
of~$E_j$ in the spectrum of the matrix $\bf S$ of $\tilde e$ is equal to one, then it
is evident that one can separate from the symmetry operator~$r^t\tilde er$ the binomial
$E_j-h$ in the power which is equal to one, wherefrom it follows that the algebraic
multiplicity of $E_j$ in the spectrum of the matrix $\bf S$ of $e$ as compared to
itself is less by two at least.  From these contradictions it follows that the
algebraic multiplicity of $E_j$ in the spectrum of the matrix $\bf S$ of $e$ is equal
to one for any~$j$. Thus, the statements 1 and 2 of Theorem~\ref{theorem3} are proved.

The statement 3 of Theorem~\ref{theorem3} can be related to the  Corollary of Lemma~1 from
\cite{ansok} if to take into account the following:

(1) one should separate intertwining operators of the f\/irst order from $e$ on
its right-hand side so that for any odd $j$ the intertwining operator, whose
matrix $\bf S$ spectrum consists of $E_j$, has as its neighbor the
intertwining operator whose matrix $\bf S$ spectrum consists of $E_{j+1}$; in
addition one must consider these pairs of neighbors as joined operators of the
second order;

(2) the properties of coef\/f\/icients in  $e$ factorization cofactors and of
corresponding intermediate Hamiltonians are easily verif\/iable by induction
from the right to the left with the help of~\eqref{v2v1},~\eqref{qpres} and the
facts that (i) Wronskian of wave functions corresponding to borders of a~forbidden band has no zeros \cite{ferneni,djferc}, (ii) the  wave function of
a Hamiltonian with periodic potential corresponding to the lower bound of the
spectrum has no zeros and (iii) an intertwining operator with periodic
coef\/f\/icients obviously maps a periodic or antiperiodic wave function (with
exception for transformation functions) to a periodic or antiperiodic wave
function respectively and increasing, decreasing or bounded Bloch
eigenfunction to increasing, decreasing or bounded Bloch eigenfunction
accordingly;

(3) there is no an intertwining operator of the second order with smooth
coef\/f\/icients with the canonical basis  of its kernel consisting of wave
functions corresponding to borders of dif\/ferent forbidden bands, this is true
because of \eqref{qpres} and due to the fact that the Wronskian of  functions
under consideration cannot be nodeless in view of dif\/ferent numbers of zeros
of these functions per period;

(4) the operator \eqref{opsim} is a $t$-antisymmetric symmetry operator for $h_i$
by virtue of the construction of Section~\ref{section3} with $h^+=h$, $h^-=h_i$,
$k_N^+=r_{(N+1)/2} \cdots  r_{i+1}$ and $p_M^+=r_1^t\cdots
r_i^t$ or $k_N^+=r_1^t\cdots r_i^t$ and
$p_M^+=r_{(N+1)/2}\cdots r_{i+1}$ depending on the relation between
orders $r_{(N+1)/2}\cdots r_{i+1}$ and $r_1^t\cdots r_i^t$
(here $N$ is not the order of $e$); non-minimizability of the operator
\eqref{opsim} follows from Theorem~\ref{theorem2}.

Theorem~\ref{theorem3} is proved.\end{proof}

\begin{corollary} \label{corollary3} Under the conditions of Theorem~{\rm \ref{theorem3}}, in view of Theorem~{\rm\ref{theorem1}} the equality
holds, \begin{gather} {\cal P}_e(h)\equiv-e^2=\prod\limits_{j=0}^{N-1}(h-E_j),\label{peh}\end{gather} and
there are $(N+1)/2$ (and not more) forbidden energy bands for the Hamiltonian $h$.
Thus, as there is a one-to-one correspondence between forbidden energy bands of $h$ and
the cofactors of an~$e$ factorization~\eqref{e32} $($see the statement $3.c$ of Theorem~{\rm\ref{theorem3}}$)$,
then the $((N+1)/2)!$ factorizations of $e$ described in Theorem~{\rm\ref{theorem3}} correspond in
one-to-one to all possible permutations of $h$ forbidden energy bands.
\end{corollary}

\begin{remark} The formula \eqref{peh} and the statement of Corollary~\ref{corollary3} about
the number of forbidden energy bands for the  periodic
solutions of stationary higher-order Korteweg -- de~Vries equations were
derived in \cite{teorsol}. The statements~1 and~2 of Theorem~\ref{theorem3} and the partial
case of the statement~3 of this theorem, corresponding to increasing of
eigenvalues of the matrices $\bf S$ for cofactors in~\eqref{e32} from the right
to the left (without formula~\eqref{opsim}) were proved in~\cite{thesis}. The
facts that the borders between allowed and forbidden bands of an arbitrary Hamiltonian $h$ with periodic potential having
$t$-antisymmetric symmetry operator $e$ correspond to certain zeros of the polynomial
${\cal P}_e(h)\equiv e^te$ and the related
wave functions belong to $\ker e$ were mentioned in~\cite{kiev}.
Special factorizati\-ons of $t$-antisymmetric non-minimizable symmetry
operator of a Hamiltonian with parity-even f\/inite-gap periodic potential can
be found in~\cite{cojapl08}.\end{remark}

\begin{remark}\label{remark4} In the case of complex periodic potential the spectrum of the matrix
$\bf S$ of $e$ can contain values located inside of the continuous spectrum of the
corresponding Hamiltonian $h$ and moreover algebraic multiplicity of these values can
be greater than one. The following example\footnote{See similar examples in
\cite{sams+}.} illustrates this situation,
\begin{gather} h=-\partial^2+{\frac{2k_0^2}{\cos^2[k_0(x-z)]}},
\qquad k_0>0,\qquad{\rm{Im}}\,z\ne0,\nonumber\\e=-p_1^-\partial p_1^+,
\qquad p_1^\mp=\pm\partial+k_0{\rm{tg}}\,[k_0(x-z)],\qquad
p_1^-=(p_1^+)^t,\nonumber\\ h_0=-\partial^2,\qquad p_1^-h_0=hp_1^-,\qquad
p_1^+h=h_0p_1^+, \label{f38}\end{gather} the eigenfunctions of $h$ continuous
spectrum $\psi_k(x)$ and  the eigenfunction of $h$ at the bottom of this
spectrum $\psi_0(x)$ take the form,
\begin{gather}\psi_k(x)=\{ ik+k_0{\rm{tg}}\,[k_0(x-z)]\}e^{ikx},\qquad
h\psi_k=k^2\psi_k, \qquad k\in\mathbb R,\nonumber\\ \psi_0(x)=k_0{\rm{tg}}\,[k_0(x-z)],
\qquad h\psi_0=0.\nonumber\end{gather} It is
interesting that there is a unique (up to a constant cofactor) bound
eigenfunction of $h$ on the level $E=k_0^2$,
\begin{gather}\psi_{0,k_0}(x)={\frac{1}{\cos[k_0(x-z)]}}\equiv-{\frac{i}{k_0}}\,e^{-ik_0z}
\psi_{k_0}(x)\equiv{\frac{i}{k_0}}\,e^{ik_0z}\psi_{-k_0}(x),\nonumber\\
h\psi_{0,k_0}=k_0^2\psi_{0,k_0},\label{f39}\end{gather} and there is a bound
associated function for this eigenfunction,
\begin{gather}\psi_{1,k_0}(x)={\frac{1}{2k_0^2}}\cos[k_0(x-z)],\qquad
(h-k_0^2)\psi_{1,k_0}=\psi_{0,k_0}.\label{f40}\end{gather} The functions
$\psi_0(x)$, $\psi_{1,k_0}(x)$ and $\psi_{0,k_0}(x)$ form a
canonical basis in the $\ker e$ by virtue of \eqref{f38}--\eqref{f40} and
\[ p_1^+\psi_{0,k_0}=0,\qquad \partial p_1^+\psi_0=0,\qquad
e\psi_{1,k_0}\equiv-p_1^-\partial p_1^+\psi_{1,k_0}=0.\] Thus, in
view of Theorem~\ref{theorem1}, \[{\cal P}_e(h)=h(h-k_0^2)^2,\qquad {\bf
S}_e=\begin{pmatrix}0&0&0\\0&k_0^2&1\\0&0&k_0^2\end{pmatrix}.\]
\end{remark}

\section[$t$-antisymmetric symmetry operators: Hamiltonians with bound state(s)]{$\boldsymbol{t}$-antisymmetric symmetry operators:\\ Hamiltonians with bound state(s)}\label{section5}

\subsection{General properties}

Let us assume that geometric multiplicity of any eigenvalue of the Hamiltonian
$h$ is~1, its algebraic multiplicity is f\/inite and functions $\psi_{l,j}(x)$
form the complete set of normalized eigenfunctions and associated functions of
$h$ for the point spectrum (without eigenvalues inside or on boundaries of
continuous spectrum),
\begin{gather} h\psi_{l,0}=E_l\psi_{l,0},\qquad (h-E_l)\psi_{l,j}=\psi_{l,j-1},\qquad
\int_{-\infty}^{+\infty}
\psi_{l,j}(x)\psi_{l',k_{l'}-j'-1}(x)\,dx=\delta_{ll'}\delta_{jj'},\nonumber\\
l,l'=0,1,2,\ldots,\qquad j=0, \ldots, k_l-1,\qquad  j'=0, \ldots,
k_{l'}-1,\label{nH}\end{gather} where $k_l$ is an algebraic multiplicity of an eigenvalue
$E_l$, $l=0, 1, 2, \dots$. It is known that in the case of Hermitian
Hamiltonian $h$ any multiplicity $k_l=1$, $l=0, 1, 2, \dots$ and it was
shown in \cite{ansok} that in this case
$e\psi_{l,0}=0$ for any $l=0, 1, 2, \dots$.

Now we derive that, in general, \begin{gather} e\psi_{l,j}=0,\qquad l=0,1,2, \dots,\quad
j=0, \dots, k_l-1.\label{enH}\end{gather} Suppose that for some $l$ there is a number
$j_0$ such that $0\leqslant j_0\leqslant k_l-1$, $e\psi_{l,j}=0$, $j=0,
\dots, j_0-1$ and $e\psi_{l,j_0}\ne0$. Then, in view of the equalities \[
he\psi_{l,j_0}=eh\psi_{l,j_0}=E_le\psi_{l,j_0}\] the function $e\psi_{l,j_0}$
is an eigenfunction of $h$ for the eigenvalue $E_l$. Hence, there is a
constant $C\ne0$ such that $e\psi_{l,j_0}=C\psi_{l,0}$. The latter leads to
contradiction by virtue of the following chain,
\begin{gather*} C=\int_{-\infty}^{+\infty}C\psi_{l,0}(x)\psi_{l,k_l-1}(x)\,dx=
\int_{-\infty}^{+\infty}[e\psi_{l,j_0}](x)\psi_{l,k_l-1}(x)\,dx\nonumber\\
\phantom{C}{} =
-\int_{-\infty}^{+\infty}\psi_{l,j_0}(x)[e\psi_{l,k_l-1}](x)\,dx=
-\int_{-\infty}^{+\infty}[(h-E_l)^{k_l-1-j_0}\psi_{l,k_l-1}](x)
[e\psi_{l,k_l-1}](x)\,dx\nonumber\\
\phantom{C}{}  =-\int_{-\infty}^{+\infty}\psi_{l,k_l-1}(x)
[e(h-E_l)^{k_l-1-j_0}\psi_{l,k_l-1}](x)\,dx\nonumber\\
\phantom{C}{} =
-\int_{-\infty}^{+\infty}\psi_{l,k_l-1}(x)
[e\psi_{l,j_0}](x)\,dx=-C,
\end{gather*} where \eqref{nH} is used. Therefore, the
equalities \eqref{enH} are valid.

It follows from \eqref{enH} that the algebraic multiplicity of $E_l$ in the
spectrum of the matrix $\bf S$ of $e$ is greater than or equal to $k_l$,
$l=0, 1, 2, \dots$  and \begin{gather}{\cal P}_e(E_l)=0\Leftrightarrow {\cal
P}_e(E_l)\psi_{l,0}={\cal P}_e(h)\psi_{l,0}=-e^2\psi_{l,0}=0,\qquad l=0,1,2, \ldots
.\label{sviaz'}\end{gather} Thus, if there is nonzero $t$-antisymmetric
non-minimizable symmetry operator $e$ for a Hamiltonian $h$, then the energies
of all its bound states satisfy the algebraic equation~\eqref{sviaz'}.

\subsection{Potentials with constant asymptotics}

The number of bound states of a Hamiltonian $h$ with nonzero $t$-antisymmetric
non\-mi\-ni\-mi\-za\-ble symmetry operator $e$ in view of \eqref{enH} and \eqref{sviaz'} is
f\/inite. Consequently, such a Hamiltonian cannot have, for example, a real-valued
potential inf\/initely increasing for $|x|\to+\infty$, because the Hamiltonians with
potentials of this type  possess (see \cite{shubin}) inf\/inite numbers of bound
states irrespectively of the rate of increasing. Taking this into account, we restrict
our consideration in this subsection by the subcase, when the potential $V(x)$ of a
Hamiltonian $h$ with a nonzero $t$-antisymmetric non\-mi\-ni\-mi\-za\-ble symmetry
operator $e$ tends to a constant $E_c$ on one of inf\/inities and either grows unboundly
 (${\rm{Re}}\,V(x)\to+\infty$, ${\rm{Im}}\,V(x)/{\rm{Re}}\,V(x)=o(1)$) or tends to a
constant dif\/ferent, in general, from $E_c$ on another inf\/inity. We assume for
def\/initeness that $V(x)\to E_c$ for $x\to-\infty$, and denote the total number of
energy levels $E_l$ as $N_b$. We shall show that under some additional assumptions of
technical character the following statements are valid.

(1) The potential $V(x)$ of the Hamiltonian $h$ is ref\/lectionless
and tends to $E_c$ for $x\to+\infty$ as well.

(2) The algebraic multiplicity of $E_l$ in the spectrum of the
matrix $\bf S$ of $e$ is equal to $2k_l$, $l=0, \ldots, N_{b}-1$.

(3) The Hamiltonian $h$ is intertwined with the Hamiltonian of a
free particle $-\partial^2+E_c$.

(4) A wave function of $h$, corresponding to the lower boundary of $h$
continuous spectrum, belongs to $\ker e$ and the energy $E_c$, corresponding
to this boundary, is contained in the spectrum of the matrix $\bf S$ of $e$
with some odd algebraic multiplicity $k_c$. Moreover, for a real-valued
potential $V(x)$ the algebraic multiplicity of $E_c$ in the spectrum of the
matrix $\bf S$ of $e$ is 1 ($k_c=1$).

(5) The spectrum of the matrix $\bf S$ of $e$ contains only $E_l$, $l=0, \ldots,
N_{b}-1$
 and $E_c$.

(6) If the order of $e$ is equal to $N$, then the number of bound states of
the Hamiltonian~$h$ is less than or equal to $(N-1)/2$. Moreover, for a
real-valued potential $V(x)$ the number of bound states of the Hamiltonian~$h$
is equal to $(N-1)/2$.

(7) For the squared symmetry operator $e$ the following
representation holds, \begin{gather} {\cal
P}_e(h)\equiv-e^2=(h-E_c)^{k_c}\prod\limits_{l=0}^{N_b-1}(h-E_l)^{2k_l}.\label{peh52}\end{gather}

(8) The operator $e$ can be represented as a product of intertwining operators
so that:

(a) \begin{gather} e=(-1)^{(N-1)/2}r_0^t\cdots r_{N_b}^t\,\partial\, r_{N_b}\cdots
r_0,\label{os77'}\\ r_l\cdots r_0\psi_{l,j}=0,\qquad l=0, \dots, N_b-1,\quad
j=0, \dots, k_l-1,\label{rl}\\ r_{N_b}\cdots r_0\psi_{c,j}=0, \qquad
j=0,\dots, \frac{k_c-3}{2},\qquad \partial\, r_{N_b}\cdots
r_0\psi_{c,(k_c-1)/2}=0,\label{c3s}\end{gather} where $\{\psi_{c,j}(x)\}_{j=0}^{k_c-1}$ is
a part of the canonical basis in $\ker e$, corresponding to the
eigenvalue~$E_c$: \[
h\psi_{c,0}=E_c\psi_{c,0},\qquad(h-E_c)\psi_{c,j}=\psi_{c,j-1},\qquad
j=0,\dots, k_c-1.\] In addition, all operators $r_0, \dots,r_{N_b}$ have
unity coef\/f\/icients at highest derivatives and other coef\/f\/icients of these
operators can have, in general, poles. Moreover, for a real-valued potential~$V(x)$,
 \begin{gather*}
  r_l = \partial+\chi_l(x),\qquad r_l\cdots
r_0\psi_{l,0} = 0  \ \Leftrightarrow \\
\qquad  \Leftrightarrow \ \chi_l(x) = -\frac{(r_{l-1}\cdots
r_0\psi_{l,0})'}{r_{l-1}\cdots r_0\psi_{l,0}},\qquad l = 0,\ldots,N_b-1,\qquad
r_{N_b}=1,
\end{gather*} all superpotentials $\chi_l(x)$, $l=0, \dots, N_b-1$ are
real-valued functions and if $V(x)\in C_{\mathbb R}^\infty$ and the energies
$E_l$, $l=0, \dots, N_b-1$ are numbered in the order of increasing, then all
these superpotentials belong to~$C^\infty_{\mathbb R}$.

(b) The intermediate Hamiltonians $h_l$, $l=1, \dots, N_b+1$,
corresponding to the fac\-to\-ri\-za\-ti\-on~\eqref{os77'}, satisfy the
following intertwinings, \begin{gather} h_l r_{l-1}=r_{l-1}h_{l-1},\qquad
r_{l-1}^th_l=h_{l-1}r_{l-1}^t,\qquad l=1,\ldots,N_b+1,\qquad
h_0\equiv h\label{int57}\end{gather} and take the Schr\"odinger form,
\begin{gather}
h_l=-\partial^2+v_l(x),\qquad l=0,\dots,N_b+1,\nonumber\\ v_0(x)=V(x),
\qquad v_{l+1}(x)=V(x)-2[\ln W_l(x)]'',\nonumber\\
W_l(x)=\begin{vmatrix}\psi_{0,0}(x)&\psi'_{0,0}(x)&\dots&
\psi_{0,0}^{(k_0+\ldots+k_l-1)}(x)\nonumber\\
\psi_{0,1}(x)&\psi'_{0,1}(x)&\dots&
\psi_{0,1}^{(k_0+\ldots+k_l-1)}(x)\nonumber\\\vdots&\vdots&\ddots&\vdots\\
\psi_{l,k_l-2}(x)&\psi'_{l,k_l-2}(x)&\dots&
\psi_{l,k_l-2}^{(k_0+\ldots+k_l-1)}(x)\\\psi_{l,k_l-1}(x)&\psi'_{l,k_l-1}(x)&\dots&
\psi_{l,k_l-1}^{(k_0+\ldots+k_l-1)}(x)\end{vmatrix},\qquad
l=0,\dots, N_b-1,\nonumber\\
W_{N_b}(x)=\begin{vmatrix}\psi_{0,0}(x)&\psi'_{0,0}(x)&\dots&
\psi_{0,0}^{(k_0+\ldots+k_{N_b-1}+(k_c-1)/2-1)}(x)\\\vdots&\vdots&\ddots&\vdots\\
\psi_{c,(k_c-3)/2}(x)&\psi'_{c,(k_c-3)/2}(x)&\dots&
\psi_{c,(k_c-3)/2}^{(k_0+\ldots+k_{N-b-1}+(k_c-1)/2-1)}(x)\end{vmatrix}.\label{hl}\end{gather}
All potentials of these Hamiltonians tend to $E_c$ for $|x|\to+\infty$ and, in
general, have poles. Moreover, in the case of real-valued potential $V(x)$ the
following chain relations take place,
\begin{gather} h_l=r_{l}^tr_{l}+E_l=r_{l-1}r_{l-1}^t+E_{l-1} ,\qquad
l=1,\ldots, N_b-1,\nonumber\\ h_0\equiv h= r_0^tr_0+E_0,\qquad
h_{N_b+1}=h_{N_b}=r_{N_b-1}r_{N_b-1}^t+E_{N_b-1},\nonumber\\
v_l(x)=\chi_{l}^2(x)-\chi'_{l}(x)+E_l=\chi_{l-1}^2(x)+\chi_{l-1}(x)+E_{l-1},\nonumber\end{gather}
all potentials $v_l(x)$, $l=0, \dots,N_b+1$ are real-valued
functions and if $V(x)\in C_{\mathbb R}^\infty$ and the energies~$E_l$,
$l=0, \dots, N_b-1$ are numbered in the order of increasing, then
all these potentials belong to $C^\infty_{\mathbb R}$.

(c) For any intermediate Hamiltonian $h_l$ there is a nonzero
$t$-antisymmetric non\-mi\-ni\-mi\-za\-ble symmetry operator $e_l$ with the
unity coef\/f\/icient at highest derivative such that:
\begin{gather*} e_lh_l=h_le_l,\qquad e_l^t=-e_l,\qquad
l=0,\ldots,{N_b+1},\qquad e_0\equiv e,\nonumber\\
e_l = (-1)^{k_l+\cdots+k_{N_b-1}+(k_c-1)/2}r_{l}^t\cdots
r_{N_b}^t\,\partial\,r_{N_b}\cdots r_{l},\qquad
l = 0,\ldots,N_b,\qquad e_{N_b+1}=\partial.
\end{gather*}

(d) The Hamiltonian $h_{N_b+1}$ is a Hamiltonian of a free particle: \[
h_{N_b+1}=-\partial^2+E_c.\]

(9) The operator $e$ acts on an eigenfunction of $h$ continuous
spectrum \begin{gather} \psi_k(x)=(-1)^{(N-1)/2}r_0^t\cdots
r_{N_b}^te^{ikx}\label{psik63}\end{gather} as follows, \begin{gather}
e\psi_k=[(ik)^{k_c}\prod_{l=0}^{N_b-1}(E_l-E_c-k^2)^{k_l}]\psi_k,\qquad
k\in\mathbb R.\label{epsik64}\end{gather}

(10) The transmission coef\/f\/icient $T(k)$ for $h$ (we assume as
usually that $T(k)$ is the ratio of the coef\/f\/icient at $e^{ikx}$ in
the main term of $\psi_k(x)$ asymptotics for $x\to+\infty$ to one
for $x\to-\infty$) takes the form\footnote{The partial case of the
formula \eqref{tr65}, corresponding to a real-valued $V(x)$, is
described in \cite{teorsol}.} \begin{gather}
T(k)=\prod_{l=0}^{N_b-1}\bigg(\frac{k+i\sqrt{E_c-E_l}}{k-i\sqrt{E_c-E_l}}\,\bigg)^{k_l},
\qquad {\rm{Re}}\,\sqrt{E_c-E_l}>0, \qquad l=0, \dots,
N_b-1.\label{tr65}
\end{gather}

In the case under consideration one can conjecture  that the coef\/f\/icients of
the operator~$e$ for $x\to-\infty$ tend to constants. If all derivatives of
the potential $V(x)$ of the Hamiltonian $h$ behave  as
$O(1/|x|^{1+\varepsilon})$, $\varepsilon>0$ for $x\to-\infty$, then the
validity of assumption on the behavior of coef\/f\/icients can be easily checked
with the help of the system of equations
\begin{gather*}\alpha_j(x)=\alpha_j(0)-\frac12[\alpha'_{j+1}(x)-\alpha'_{j+1}(0)]-
\frac12\int_0^x\sum\limits_{l=j+2}^NC_l^{j+2}\alpha_l(t)V^{(l-j-1)}(t)\,dt,\nonumber\\
 j=N-2, \ldots, 0
 \end{gather*} with respect to the coef\/f\/icients $\alpha_j(x)$, $j=0$,
\dots, $N$ of the operator\footnote{The identity $\alpha_{N-1}(x)\equiv0$ is a
consequence of the identity $\alpha_N(x)\equiv1$ and of $t$-antisymmetry of
the ope\-ra\-tor~$e$.} \[
e=\sum\limits_{j=0}^N\alpha_j(x)\partial^j,\qquad\alpha_N(x)\equiv1,
\qquad\alpha_{N-1}(x)\equiv0.\] This system follows from the condition
$eh=he$.

First, let us show that the spectrum of the matrix $\bf S$ of $e$ contains
only the values $E_l$, $l=0$, $\dots, N_b-1$, $E_c$ and (in the case of
existence of a f\/inite limit $E'_c$ of $V(x)$ for $x\to+\infty$) the value~$E'_c$. Belonging of $E_l$, $l=0, \dots, N_b$ to this spectrum was derived
before. For the values of spectral parameter $\lambda$ such that
$\lambda-E_c\geqslant0$ there are formal eigenfunctions of~$h$, which for
$x\to-\infty$ are proportional to~$e^{ikx}$ and to~$e^{-ikx}$,
$k=\sqrt{\lambda-E_c}$. The formal eigenfunctions of~$h$,
 proportional to~$e^{ikx}$ and to $e^{-ikx}$ for $x\to-\infty$,  are formal eigenfunctions of
$e$ by virtue of constant asymptotics of $e$ coef\/f\/icients. In view of
$t$-antisymmetry of $e$ , the related eigenvalues take the form
$kf(k^2)$ and $-kf(k^2)$ correspondingly  where $f(k^2)$ is a certain
function. In addition, $f(k^2)$ cannot have zeros for real $k\ne0$, since in
the opposite case $\ker e$ contains two linearly independent formal
eigenfunctions of $h$ for the same value of a spectral parameter, that
contradicts to non-minimizability of $e$  in view of Theorem~\ref{theorem2}. Thus, the
spectrum of the matrix $\bf S$ of $e$ contains~$E_c$ and cannot include
$\lambda$, which satisfy $\lambda-E_c>0$. An analogous statement is valid also
for f\/inite $E'_c$. For any spectral value $\lambda$, which does not satisfy
$\lambda-E_c\geqslant0$ and $\lambda-E'_c\geqslant0$ and is dif\/ferent from
energies $E_l$, $l=0, \dots, N_b-1$, there is~\cite{andcansok07,sok1} a
formal eigenfunction of $h$, which tends to zero for $x\to-\infty$, and a
formal eigenfunction of $h$, which tends to zero for $x\to+\infty$. These
eigenfunctions are evidently linearly independent formal eigenfunctions of $e$
and squared corresponding eigenvalues by virtue of~\eqref{harpol} are equal to
$-{\cal P}_e(\lambda)$. If a considered $\lambda$ belongs to the spectrum of
the matrix $\bf S$ of $e$, then ${\cal P}_e(\lambda)=0$ and the eigenfunctions
mentioned  above belong to $\ker e$, then it contradicts to non-minimizability
of $e$  in view of Theorem~\ref{theorem2}. Thus, the spectrum of the matrix $\bf S$ of $e$
contains only the values $E_l$, $l=0, \dots, N_b-1$, $E_c$ and (in the case
of existence of a f\/inite limit $E'_c$ of $V(x)$ for $x\to+\infty$) the value
$E'_c$.

Now we derive that the algebraic multiplicity of any energy $E_l$, $l=0,\dots,N_b-1$ in the spectrum of the matrix $\bf S$ of $e$ is equal $2k_l$.
With the help of Lemma 1 from \cite{ansok} one can represent the operator $e$
so that \begin{gather*}
 e=\hat er_0,\qquad r_0\psi_{0,j}=0,\qquad j=0,\ldots,k_0-1,\\
\hat eh_1=h\hat e,\qquad r_0h=h_1r_0,\qquad hr_0^t=r_0^th_1,
\end{gather*}
where $r_0$ and $h_1$ are def\/ined in \eqref{rl}, \eqref{hl} and the coef\/f\/icients of
$r_0$, $\hat e$ and the potential of $h_1$ have poles in general. In
accordance to Corollary~2 from \cite{sok1} the main terms of asymptotics for
$x\to+\infty$ ($x\to-\infty$) of the potentials in the Hamiltonians $h\equiv
h_0$ and $h_1$ are identical.

One can continue $h$, $h_1$, $r_0$ and $\hat e$ for $x$, to a some path in
complex plain, which avoids all above mentioned poles and can be identif\/ied to
real axis for suf\/f\/iciently large $|x|$ (absence of real poles for large $x$
follows from invariance of the class $\cal K$ with respect to intertwinings
proved in~\cite{sok1}). Using this conjecture and arguments analogous to ones
in the proof of the index theorem in \cite{sok1}, we can derive that this
theorem is valid for the case under consideration and thereby the Hamiltonian
$h_1$ does not possess normalizable eigenfunctions and associated functions
for the spectral value $E_0$. Hence, the functions $\psi_{0,j}$, $j=0, \dots,
k_0-1$ belong to~$\ker \hat e^t$ since in the opposite case the intertwining
operator~$\hat e^t$ maps these functions into the chain of eigenfunction and
associated functions of~$h_1$ for the eigenvalue~$E_0$.

With the help of Lemma 1 from \cite{ansok} we can represent the operator $\hat
e^t$ in the form \[ \hat e^t= (-1)^{k_0}e_1^tr_0,\qquad e_1h_1=h_1e_1,\qquad
e_1^t h_1=h_1 e_1^t,\] where the coef\/f\/icient at the highest derivative in $e_1$ is
equal to 1. Thus, the symmetry operator $e$ can be obviously factorized in the
form \[ e=(-1)^{k_0}r_0^te_1r_0 \] and the symmetry operator $e_1$ is
non-minimizable and $t$-antisymmetric, because otherwise the operator $e$ is
minimizable and/or $e+e^t=(-1)^{k_0}r_0^t(e_1+e_1^t)r_0\ne0$. Taking into
account that in view of Theorem~\ref{theorem1} the spectrum of a product of intertwining
operators is equal to  a union  of the spectra of the matrices $\bf S$ of the
cofactors (with regard to algebraic multiplicities) and that the spectrum of
the matrix $\bf S$ of $e_1$ does not contain $E_0$, we derive that the algebraic
multiplicity of~$E_0$ for the spectrum of the matrix $\bf S$ of $e$ is equal
to $2k_0$.

Using the fact, that the eigenfunctions and associated functions of $h_1$ and
the corresponding  eigenvalues take the form $r_0\psi_{l,j}(x)$, $j=0, \dots,
k_l-1$ and $E_l$ respectively, $l=1, \dots, N_b-1$, and inductive reasoning
as well, we conclude that the algebraic multiplicity of $E_l$ in the spectrum
of the matrix $\bf S$ of $e$ is equal to $2k_l$, $l=0, \dots, N_b-1$ and that
the corresponding part of the statement (8) is valid. The related part of the
statement~(8) for the case of real-valued $V(x)$ and for the subcase $V(x)\in
C^\infty_{\mathbb R}$ is evidently valid in virtue of Lemma 1 from~\cite{ansok} and due
to the fact that eigenfunctions of $h$ in this case can be chosen real-valued
and that the eigenfunction of a Hermitian Hamiltonian for its ground state does
not have zeros.

Let us check now that the potential of $h$  tend to $E_c$ for $x\to+\infty$ as well as the potentials
of all intermediate Hamiltonians. In so far as
the order of the operator $e$ is odd, the sum of algebraic
multiplicities of~$E_c$ and (if $V(x)\to E'_c$, $x\to+\infty$)
$E'_c$ in the spectrum of the matrix~$\bf S$ of~$e$ is obviously odd
as well. Hence the algebraic multiplicity of either~$E_c$ or~$E'_c$
is odd. We shall restrict ourselves by the case, when the
multiplicity of $E_c$ is odd, because the examination of the
opposite case is analogous.

With the help of Lemma 1 from \cite{ansok} one can factorize the symmetry
operator $e_{N_b}$ in the product of intertwining operators of the f\/irst
order so that f\/irst $k'_c/2$ and last $k'_c/2$ operators in this
factorization correspond to the eigenvalue $E'_c$ of the matrix $\bf S$ of
$e_{N_b}$, where $k'_c$ is an algebraic multiplicity of $E'_c$. This
factorization is unique, since for any step of this factorization only the
unique (up to constant cofactor) eigenfunction of corresponding intermediate
Hamiltonian from the kernel of factorized operator can form a basis in the
kernel of a separated intertwining operator of the f\/irst order.
From uniqueness of considered factorization and from $t$-antisymmetry of
$e_{N_b}$ it follows that the central place in this factorization (i.e.\  the
$(N-2k_0-\cdots-2k_{N_b-1}+1)/2\equiv(k_c+k'_c+1)/2$-th position from the right or from
the left) is occupied by~$\partial$ and that the operator~$e_{N_b}$ can be
represented in the form \[ e_{N_b}=(-1)^{(k_c+k'_c-1)/2}r^t\,\partial\, r,\]
where $r$ is intertwining operator of the $(k_c+k'_c-1)/2$-th order. In
addition, according to Lemma~1 from \cite{ansok}, the operator $r$ intertwines
the Hamiltonian $h_{N_b}$ with the Hamiltonian \[
h_{N_b+1}=\partial^t\partial+E_c\equiv-\partial^2+E_c,\] i.e.\  with the
Hamiltonian of a free particle.

If $V(x)\to E_c$ for $x\to+\infty$, obviously $r$ is
identical to $r_{N_b}$, def\/ined in \eqref{c3s}, and the potentials\ of
all intermediate Hamiltonians tend to $E_c$ for $|x|\to+\infty$. If $V(x)$ inf\/initely increases
(${\rm{Re}}\,V(x)\to+\infty$,
${\rm{Im}}\,V(x)/{\rm{Re}}\,V(x)=o(1)$) for $x\to+\infty$, the
operator $r$ is equal to $r_{N_b}$ as well and the potentials of the
intermediate Hamiltonians $h_1, \dots, h_{N_b}$ inf\/initely
increase  for $x\to+\infty$ as well. On the other hand, a canonical
basis in $\ker r^t=\ker r_{N_b}^t$ consists of the chain of an
eigenfunction and associated functions of $h_{N_b+1}$ for the
spectral value $E_c$ and all these functions are evidently polynomials.
Hence the Wronskian of these functions is a polynomial
as well and in view of~\eqref{v2v1} the potential in~$h_{N_b}$  tends to
$E_c$ for $x\to +\infty$, that contradicts to what has been stated above.
Consequently the potential $V(x)$ cannot increase unboundly
for $x\to+\infty$.

Now we analyze the case, when $V(x)$ tends to a f\/inite constant
$E'_c\ne E_c$ for $x\to+\infty$. In this case the potentials of the
intermediate Hamiltonians $h_1, \dots, h_{N_b}$ tend to $E'_c$ for
$x\to+\infty$ as well. In accordance to the  factorization mentioned
above the operator $r$  can be represented as follows, \[ r=r^{(a)}r^{(b)},\]
where $r^{(a)}$ and $r^{(b)}$ are intertwining operators of the orders
$(k_c-1)/2$ and $k'_c/2$ respectively, all eigenvalues of the matrix
$\bf S$ of $r^{(a)}$ are equal $E_c$ and all eigenvalues of the matrix
$\bf S$ of $r^{(b)}$ are equal $E'_c$. If
$E'_c-E_c\in{\mathbb C}\setminus[0,+\infty)$ the potential of the
Hamiltonian $h'$, intertwined by $r^{(a)}$ with $h_{N_b+1}$
 tends to $E_c$ for $x\to+\infty$ and the
potential of the Hamiltonian $h_{N_b}$, intertwined by $r^{(b)}$ with
$h'$, tends to $E_c$ for
$x\to+\infty$  by virtue of Corollary 2 from \cite{sok1}. The latter contradicts to what has been written above
and therefore $E'_c-E_c>0$.

Let us re-factorize the operator $r$ with the help of Lemma 1 from
\cite{ansok} in the form \begin{gather*}
r=r^{(c)}r^{(d)},\end{gather*} where $r^{(c)}$ and $r^{(d)}$ are
intertwining operators of the orders $k'_c/2$ and $(k_c-1)/2$ respectively,
all eigenvalues of the matrix $\bf S$ of $r^{(c)}$ are equal $E'_c$ and all
eigenvalues of the matrix $\bf S$ of $r^{(d)}$ are equal $E_c$. The potential of
the intermediate Hamiltonian $h''$, intertwined by $r^{(d)}$ with $h_{N_b}$,
 tends to $E'_c$ for $x\to+\infty$ according to Corollary 2 from \cite{sok1}.
Thus, the Wronskian $W(x)$ of elements of a basis in $\ker(r^{(c)})^t$, in view of
\eqref{v2v1}, can be estimated in the following way: \begin{gather} [\ln
W(x)]''=-\tfrac12\,{k'}^2+o(1)\Rightarrow \ln
W(x)=-\tfrac14\,{k'}^2x^2+o(x^2)\Rightarrow
W(x)=e^{-k^{\prime2}x^2/4+o(x^2)},\nonumber\\ x\to+\infty,\qquad
k'=\sqrt{E'_c-E_c}>0.\label{wr72}\end{gather} On the other hand, a canonical basis in
$\ker (r^{(c)})^t$ consists of a chain of eigenfunction and associa\-ted functions
of $h_{N_b+1}$ for the spectral value $E'_c$ and all these functions are
linear combinations of $e^{ik'x}$ and $e^{-ik'x}$ with polynomial
coef\/f\/icients. Their Wronskian  obviously cannot be of the form~\eqref{wr72}. Thus, the inequality $E'_c-E_c>0$ cannot be realized also and the potential
$V(x)$ as well as the potentials of all intermediate Hamiltonians tend to
$E_c$ for $|x|\to+\infty$.

It was noticed above, that the formal eigenfunctions of $h$
proportional to $e^{ikx}$ and to $e^{-ikx}$ for $x\to-\infty$ are
formal eigenfunctions of $e$, and corresponding eigenvalues take the
form $kf(k^2)$ and $-kf(k^2)$ respectively, where $f(k^2)\ne0$ for
real $k\ne0$. The same obviously takes place  for $x\to+\infty$ as well.
Moreover, it is evident, that the linear combination of these functions
with nonzero coef\/f\/icients cannot be an eigenfunction of $e$.
Hence, the potential $V(x)$ is ref\/lectionless, unless an  eigenfunction of $h$ is proportional to $e^{ikx}$
for $x\to+\infty$ and to $e^{-ikx}$ for $x\to-\infty$ (or
respectively to $e^{-ikx}$ and to $e^{ikx}$). The latter is
impossible in view of \eqref{psik63} and of constant asymptotics of $r_0,
\dots, r_{N_b}$ coef\/f\/icients, which follows from the fact, that:

(1) the operator $r_l$ for $x\to\pm\infty$ is asymptotically equal
to \begin{gather}(\partial\pm\sqrt{E_c-E_l}\,)^{k_l}, \qquad
{\rm{Re}}\,\sqrt{E_c-E_l}>0,\qquad l=0, \dots, N_b-1,\label{as1}\end{gather}
because  an element of the kernel of any cofactor of $r_l$
factorization, obtained in accordance to Lemma 1 from \cite{ansok}, is proportional
to $e^{\mp\sqrt{E_c-E_l}\,x}$ for $x\to\pm\infty$ (being an eigenfunction of the corresponding intermediate
Hamiltonian);

(2) the operator $r_{N_b}$ for $x\to\pm\infty$ is asymptotically
equal to \begin{gather}\partial^{(k_c-1)/2},\label{as2}\end{gather} because, as was noticed
above, the canonical basis in $\ker r^t_{N_b}$ consists of
polynomials and consequently an element of the kernel of any
cofactor of $r^t_{N_b}$ factorization, obtained in accordance to
Lemma 1 from \cite{ansok}, is a rational function.

The formula \eqref{epsik64} is a consequence of
\eqref{os77'}, \eqref{int57}, \eqref{psik63} and Theorem~\ref{theorem1}. The
representation \eqref{tr65} for $T(k)$ follows from \eqref{psik63},
\eqref{as1} and \eqref{as2}.

At last we derive that $r_{N_b}=1$ for real-valued
$V(x)$. Let us assume the opposite and demonstrate, that this tends
to a contradiction. For this purpose we show at f\/irst, that the
Wronskians of elements of a canonical basis $\varphi_1(x), \dots,
\varphi_{(k_c-1)/2}$ in $\ker r^t_{N_b}$ satisfy the following
system, \begin{gather} \Big(\frac{\hat W_l(x)}{\hat
W_{l-2}(x)}\Big)'=-\Big({\frac{\hat W_{l-1}(x)}{\hat
W_{l-2}(x)}}\Big)^2,\qquad l=2, \ldots, \frac{k_c-1}{2},\nonumber\\ \hat
W_0(x)\equiv1,\qquad\hat
W_l(x)=\begin{vmatrix}\varphi_1(x)&\varphi'_1(x)&\ldots&\varphi_1^{(l-1)}(x)\\
\vdots&\vdots&\ddots&\vdots\\
\varphi_l(x)&\varphi'_l(x)&\ldots&\varphi_l^{(l-1)}(x)\end{vmatrix},\qquad
l=1, \ldots, \frac{k_c-1}{2}. \label{sys75}\end{gather} This system arises by virtue
of Lemma 1 from \cite{ansok}, when using the factorization $r^t_{N_b}$ in the product
of intertwining operators of the f\/irst order  and owing to
\eqref{qpres},
\begin{gather} r^t_{N_b}=(-1)^{(k_c-1)/2}\hat r_{N_b}\cdots\hat r_1,\qquad \hat r_l\cdots\hat
r_1\varphi_l=0,\qquad\hat
r_l=\partial+\hat\chi_l(x),\nonumber\\
\hat\chi_l(x)= -{\frac{(\hat
r_{l-1}\cdots\hat r_1\varphi_l(x))}{\hat r_{l-1}\cdots\hat
r_1\varphi_l(x)}}'=-{\frac{(\hat W_l(x)/\hat W_{l-1}(x))}{\hat
W_l(x)/\hat W_{l-1}(x)}}'={\frac{\hat W'_{l-1}(x)}{\hat
W_{l-1}(x)}}-{\frac{\hat W'_{l}(x)}{\hat W_{l}(x)}},\nonumber\\
l=1,\ldots, \frac{k_c-1}{2}\nonumber
\end{gather} and employing the chain:
\begin{gather}
\left({\frac{\hat W_{l-1}(x)}{\hat W_{l-2}(x)}}\right)^2={\frac{\hat
W_{l-1}(x)}{\hat W_{l-2}(x)}}\hat r_{l-2}\cdots\hat
r_1\varphi_{l-1}={\frac{\hat W_{l-1}(x)}{\hat W_{l-2}(x)}}\hat
r_{l-2}\cdots\hat r_1(h_{N_b+1}-E_c)\varphi_{l}\nonumber\\
\phantom{\left({\frac{\hat W_{l-1}(x)}{\hat W_{l-2}(x)}}\right)^2}{} ={\frac{\hat W_{l-1}(x)}{\hat W_{l-2}(x)}}(\hat h_{l-2}-E_c)\hat
r_{l-2}\cdots\hat r_1\varphi_{l}={\frac{\hat W_{l-1}(x)}{\hat
W_{l-2}(x)}}\hat r^t_{l-1}\hat r_{l-1}\hat r_{l-2}\cdots\hat
r_1\varphi_{l}\nonumber\\
\phantom{\left({\frac{\hat W_{l-1}(x)}{\hat W_{l-2}(x)}}\right)^2}{}
={\frac{\hat W_{l-1}(x)}{\hat W_{l-2}(x)}}
\left(-\partial+{\frac{\hat W'_{l-2}(x)}{\hat W_{l-2}(x)}}-{\frac{\hat
W'_{l-1}(x)}{\hat W_{l-1}(x)}}\right){\frac{\hat W_{l}(x)}{\hat
W_{l-1}(x)}}\nonumber\\
\phantom{\left({\frac{\hat W_{l-1}(x)}{\hat W_{l-2}(x)}}\right)^2}{}
=-\left({\frac{\hat W_l(x)}{\hat W_{l-2}(x)}}\right)',\quad
l=2, \ldots, \frac{k_c-1}{2},\nonumber
\end{gather} where $\hat h_1, \dots, \hat h_{(k_c-5)/2}$ are the corresponding intermediate Hamiltonians and
$\hat h_0=h_{N_b+1}$.

With the help of the system \eqref{sys75} we shall demonstrate, that it is
possible to separate from the right-hand side of $r^t_{N_b}$ the intertwining operator
of the f\/irst or  of the second order with smooth coef\/f\/icients. For this purpose, in view of \eqref{v2v1} and \eqref{qpres},
it is suf\/f\/icient  to derive, that there are
no zeros either for $\hat W_2(x)$ or for $\hat W_{(k_c-1)/2-2l}(x)$, $l=1, 2,3, \dots$. Assume, that there are zeros for both $\hat W_2(x)$ and $\hat
W_{(k_c-1)/2-2}(x)$, and show, that this assumption is contradictory.
Let us  notice, that the Wronskian $\hat
W_{(k_c-1)/2}(x)$ has no zeros by virtue of \eqref{v2v1} and of inf\/inite smoothness
of the potentials of $h_{N_b}$ and $h_{N_b+1}$ and that all functions
$\varphi_1(x), \dots, \varphi_{(k_c-1)/2}(x)$ being polynomials
possess f\/inite numbers of zeros. Assume also without loss of
generality that all these functions are real-valued.

{\sloppy
In view of \eqref{sys75} the ratio $\hat W_{(k_c-1)/2}(x)/\hat
W_{(k_c-2)/2-2}(x)$ decreases monotonically from $+\infty$ starting
from the utmost right zero of $\hat W_{(k_c-2)/2-2}(x)$ and tends
to a nonnegative limit for $x\to+\infty$. Using equation~\eqref{sys75} for
two successive $l$, one can obtain the system \begin{gather} \left(-{\frac{\hat
W_{l}(x)}{\hat W_{l-2}(x)}}\right)'\left({\frac{\hat W_{l-3}(x)}{\hat
W_{l-1}(x)}}\right)'=1,\qquad l=3, \ldots, \frac{k_c-1}{2}. \label{83}
\end{gather} With the help of the equation from this system for $l=(k_c-1)/2$
we conclude, that the ratio ${{\hat W_{(k_c-1)/2-3}(x)}/{\hat
W_{(k_c-1)/2-1}(x)}}$ monotonically increases towards the right
side, starting from the utmost right zero of $\hat
W_{(k_c-1)/2-1}(x)$, $\hat W_{(k_c-1)/2-2}(x)$  and $\hat
W_{(k_c-1)/2-3}(x)$.
As well, in view of equation~\eqref{83}
 and the Bunyakovsky inequality, the following estimate holds,
\begin{gather}              (x-x_0)^2\leqslant\left({\frac{\hat W_{(k_c-1)/2}(x_0)}{\hat
W_{(k_c-1)/2-2}(x_0)}}-{\frac{\hat W_{(k_c-1)/2}(x)}{\hat
W_{(k_c-1)/2-2}(x)}}\right)\left({\frac{\hat W_{(k_c-1)/2-3}(x)}{\hat
W_{(k_c-1)/2-1}(x)}}-{\frac{\hat W_{(k_c-1)/2-3}(x_0)}{\hat
W_{(k_c-1)/2-1}(x_0)}}\right),\!\nonumber\\ x>x_0,\label{bun79}
\end{gather} where $x_0$
is a f\/ixed point on the right-hand side of the utmost right zero of
$\hat W_{(k_c-1)/2-1}(x)$,  $\hat W_{(k_c-1)/2-2}(x)$ and
$\hat W_{(k_c-1)/2-3}(x)$. The left-hand side of \eqref{bun79} tends to
$+\infty$ for~\mbox{$x\to+\infty$} and the f\/irst cofactor on the right-hand
side of \eqref{bun79} approaches to a positive constant for~\mbox{$x\to+\infty$}. Hence, the ratio $\hat W_{(k_c-1)/2-3}(x)/\hat
W_{(k_c-1)/2-1}(x)$ tends to $+\infty$ for \mbox{$x\to+\infty$} and $\hat
W_{(k_c-1)/2-1}(x)/\hat W_{(k_c-1)/2-3}(x)$ monotonically decreases
for $x>x_0$ and tends to zero for~\mbox{$x\to+\infty$}. Arguing in the same
way by induction, one obtains that the ratios $\hat
W_{(k_c-1)/2-l-3}(x)/\hat W_{(k_c-1)/2-l-1}(x)$, $l=1, \dots,
(k_c-1)/2-3$ also tend to $+\infty$ for $x\to+\infty$, but the
latter contradicts to the fact, that $\hat W_2(x)\equiv\hat
W_2(x)/\hat W_0(x)$  monotonically decreases on the whole axis in
view of~\eqref{sys75} and is negative on the right from the unique zero
of $\hat W_2(x)$. Thus, there are no zeros, at least, for one of the
Wronskians $\hat W_2(x)$ and $\hat W_{(k_c-1)/2-2}(x)$. Hence, in
view of~\eqref{v2v1} and~\eqref{qpres} one can separate an intertwining
operator of the second order with inf\/initely smooth coef\/f\/icients
from one of the sides of $r^t_{N_b}$. Using induction again, we
conclude that it is possible to separate the intertwining operator
of the f\/irst or of the second order with inf\/initely smooth
coef\/f\/icients from the right-hand side of~$r^t_{N_b}$.

}

Finally let us demonstrate, that the latter result tends to a contradiction. The
function $\varphi_1(x)$ as a formal eigenfunction of $h_{N_b+1}$ for the
spectral value $E_c$ takes either the form $\varphi_1(x)=C$, $C\in\mathbb R$,
$C\ne0$ or the form $\varphi_1(x)=C_1x+C_2$, $C_1\in\mathbb R$, $C_2\in
\mathbb R$, $C_1\ne0$. The f\/irst one is impossible, because in this
case $\hat r_1=\partial$, that contradicts to non-minimizability of $e$ ($\hat
r_1\partial=\partial^2=E_c-h_{N_b+1}$). For the second one the  separation in
$r^t_{N_b}$ on its right-hand side of the intertwining operator of the  f\/irst
order with inf\/initely smooth coef\/f\/icients is impossible because the coef\/f\/icient
of $\hat r_1=\partial-C_1/(C_1x+C_2)$ at $\partial^0$ possess the pole at
$x=-C_2/C_1$. The separation of the intertwining operator of the
second order in $r^t_{N_b}$ on the right-hand side is impossible as well,
because one can easily check that $\hat W_2(x)$ cannot be nodeless.
Therefore, $r_{N_b}=1$.

Thus, all statements (1)--(9) from the beginning of this subsection
are validated.

\begin{remark} It follows from \eqref{tr65}, that in the presence of non-real
energy(-ies) of $h$ bound state(s) the value $|T(k)|$ is dif\/ferent from
identical unity. But if all non-real energies among $E_l$, $l=0$, \dots,
$N_b-1$ can be divided into pairs of mutually complex conjugated energies with
equal (inside a pair) algebraic multiplicities, then obviously
$|T(k)|\equiv1$.\end{remark}

\section{Examples}\label{section6}

We present here three examples\footnote{Other relevant examples can found in
\cite{sams+}.}, illustrating results of the previous
section.
\begin{example} Non-Hermitian (in general) Hamiltonian with one
bound state
\begin{gather*} h=-\partial^2-{\frac{2\alpha^2}{{\rm ch}^2\alpha x}},\qquad
{\rm{Re}}\,\alpha>0,\nonumber\\ \psi_{0,0}(x)=\frac{1}{{\rm ch}\,\alpha
x},\qquad h\psi_{0,0}=E_0\psi_{0,0},\qquad E_0=-\alpha^2,\nonumber\\
\psi_c(x)={\rm th}\,\alpha x,\qquad h\psi_c=E_c\psi_c,\qquad
E_c=0,\nonumber\\ e=-r_0^t\,\partial\,r_0,\nonumber\\
r_0=\partial-{\frac{\psi^\prime_{0,0}(x)}{\psi_{0,0}(x)}}\equiv\partial+\alpha\,{\rm
th}\,\alpha x,\qquad r_0\psi_{0,0}=0,\qquad \partial r_0\psi_c=0,\nonumber\\
{\bf
S}_e=\begin{pmatrix}E_c&0&0\\0&E_0&1\\0&0&E_0\end{pmatrix},\nonumber\\
{\cal P}_e(h)=(h-E_c)(h-E_0)^2,\nonumber\\\psi_k(x)=(ik-\alpha\,{\rm
th}\,\alpha x)e^{ikx},\qquad h\psi_k=k^2\psi_k,\qquad
T(k)={\frac{k+i\alpha}{k-i\alpha}},\qquad k\in \mathbb R.
\end{gather*}
\end{example}

\begin{example} Non-Hermitian Hamiltonian with one Jordan cell of the
2nd order
\begin{gather*} h=-\partial^2-16\alpha^2{\frac{\alpha(x-z){\rm sh}\,2\alpha x-2{\rm
ch}^2\alpha x}{[{\rm sh}\,2\alpha x+2\alpha(x-z)]^2}},\qquad
\alpha>0,\qquad {\rm Im}\,z\ne0,\nonumber\\ \psi_{0,0}(x)={\frac{{\rm
ch}\,\alpha x}{{\rm sh}\,2\alpha x+2\alpha(x-z)}},\qquad
\psi_{0,1}(x)={\frac{2\alpha(x-z){\rm sh}\,2\alpha x-{\rm ch}\,\alpha
x}{(2\alpha)^2[{\rm sh}\,2\alpha x+2\alpha(x-z)]^2}},\nonumber\\
h\psi_{0,0}=E_0\psi_{0,0},\qquad(h-E_0)\psi_{0,1}=\psi_{0,0},\qquad
E_0=-\alpha^2,\nonumber\\ \psi_c(x)={\frac{{\rm sh}\,2\alpha
x-2\alpha(x-z)}{{\rm sh}\,2\alpha x+2\alpha(x-z)}},\qquad
h\psi_c=E_c\psi_c,\qquad E_c=0,\nonumber\\ e=r_0^t\,\partial\,
r_0,\qquad r_0=r_{0,1}r_{0,0},\nonumber\\
r_{0,0}=\partial-{\frac{\psi^\prime_{0,0}(x)}{\psi_{0,0}(x)}},\qquad
r_{0,1}=\partial-{\frac{(r_{0,0}\psi_{0,1})^\prime(x)}{r_{0,0}\psi_{0,1}(x)}},\nonumber\\
 r_{0,0}\psi_{0,0}=0,\qquad r_{0,1}r_{0,0}\psi_{0,1}=0,\qquad
\partial r_0\psi_c=0,\nonumber\\{\bf
S}_e=\begin{pmatrix}E_c&0&0&0&0\\0&E_0&1&0&0\\0&0&E_0&1&0\\0&0&0&E_0&1\\0&0&0&0&E_0
\end{pmatrix},\nonumber\\
 {\cal P}_e(h)=(h-E_c)(h-E_0)^4,\nonumber\\
\psi_k(x)={\frac{(\alpha^2-k^2){\rm sh}\,2\alpha x-2i\alpha k(1+{\rm
ch}\,2\alpha x)-2\alpha(\alpha^2+k^2)(x-z)}{{\rm sh}\,2\alpha
x+2\alpha(x-z)}}\,e^{ikx},\nonumber\\  h\psi_k=k^2\psi_k,\qquad
T(k)=\Big({\frac{k+i\alpha}{k-i\alpha}}\Big)^2,\qquad k\in\mathbb R.
\end{gather*}
\end{example}

\begin{example} Non-Hermitian Hamiltonian with one bound state at
the bottom of continuous spectrum
\begin{gather} h=-\partial^2+{\frac{2}{(x-z)^2}},\qquad {\rm Im}\,z\ne0,\nonumber\\
\psi_{c,0}(x)=\frac{1}{(x-z)}\in L_2({\mathbb R}),\qquad
h\psi_{c,0}=E_c\psi_{c,0},\qquad E_c=0,\nonumber\\
\psi_{c,1}(x)=\frac12(x-z),\qquad(h-E_c)\psi_{c,1}=\psi_{c,0},\nonumber\\
 e=-r_0^t\,\partial\,r_0,\nonumber\\
r_0=\partial-{\frac{\psi^\prime_{c,0}(x)}{\psi_{c,0}(x)}}\equiv\partial+\frac{1}{x-z},\qquad
r_0\psi_{c,0}=0,\qquad \partial r_0\psi_{c,1}=0,\nonumber\\
{\bf
S}_e=\begin{pmatrix}E_c&1&0\\0&E_c&1\\0&0&E_c\end{pmatrix},\nonumber\\
{\cal P}_e(h)=(h-E_c)^3,\label{peh109}\\
\psi_k(x)=\Big(ik-\frac{1}{x-z}\Big)e^{ikx},\qquad
h\psi_k=k^2\psi_k,\qquad T(k)=1,\qquad k\in \mathbb R.\nonumber
\end{gather}
\end{example}

\section{Concluding remarks and generalizations}\label{section7}

(1) Let us examine the situation when the Hamiltonian $h^+=-\partial^2+V_1(x)$ with a smooth
real-valued periodic potential $V_1(x)$ is transformed into the
Hamiltonian $h^-=-\partial^2+V_2(x)$ with a~smooth real-valued
potential $V_2(x)$, whose spectrum is dif\/ferent from the spectrum of~$h^+$ only by presence of an eigenvalue $\lambda_1$ or two eigenvalues $\lambda_1$ and $\lambda_2$.  The former can be done~\mbox{\cite{ferneni,djferc}} with the help of one
nodeless real-valued transformation function $\phi_1(x)$, which is
non-Bloch formal eigenfunction of~$h^+$ for a real spectral value
$\lambda_1$, situated below continuous spectrum of~$h^+$.  The latter can be realized with the help of two real-valued
transformation functions~$\phi_1(x)$ and~$\phi_2(x)$ with nodeless
Wronskian $W_-(x)= \phi_1^-(x)\phi_2^{-\prime}(x) -
\phi_1^{-\prime}(x)\phi_2^-(x)$, which are non-Bloch formal
eigenfunctions of $h^+$ for a real spectral values $\lambda_1$ and
$\lambda_2\ne\lambda_1$ respectively, situated inside a forbidden
energy band (the same for both values) of~$h^+$.

Let us suppose that for $h^+$ there is a non-minimizable
$t$-antisymmetric symmetry operator~$e^+$ with unity coef\/f\/icient at the
derivative of the highest order, \[ h^+e^+=e^+h^+,\qquad (e^+)^t=-e^+,\] and
that $q_1^\pm$ ($q_2^\pm$) are corresponding intertwining operators for the
f\/irst (second) case mentioned above, \begin{gather*} h^\pm q_1^\pm=q_1^\pm h^\mp,\qquad
(q_1^+)^t=q_1^-,\qquad q_1^-\phi_1^-=0,\nonumber\\ h^\pm q_2^\pm=q_2^\pm
h^\mp,\qquad (q_2^+)^t=q_2^-,\qquad q_2^-\phi_1^-=q_2^-\phi^-_2=0.
\end{gather*} Then,
for $h^-$ there is obviously a nonzero $t$-antisymmetric symmetry operator $e^-$
with the unity coef\/f\/icient at the highest derivative: \begin{gather} h^-e^-=e^-h^-,\qquad
(e^-)^t=-e^-,\qquad e^-=(-1)^j(q_j^+)^te^+q_j^+,\qquad j=1,2.\label{e114}\end{gather}
Moreover, in view of \eqref{enH} and Theorems~\ref{theorem2} and~\ref{theorem3} the operator $e^-$ is
non-minimizable and the canonical basis in $\ker q_1^+$ ($\ker q_2^+$) consists
of eigenfunction(s) of $h^-$ for the eigenvalue(s) $\lambda_1$ ($\lambda_1$~and~$\lambda_2$). As well by virtue of~\eqref{v2v1}, \eqref{qpres}
and Theorem~\ref{theorem3}  one can conclude  that if $V_1(x)\in C^\infty_\mathbb R$, $l=1, 2$ the potential
$V_2(x)$ and coef\/f\/icients of $q_j^\pm$ and $e^-$ are inf\/initely smooth too.

It follows from Theorem~\ref{theorem1} and \eqref{e114}, that  \begin{gather}{\cal
P}_{e^-}(h^-)\equiv-(e^-)^2={\cal
P}_{e^+}(h^-)\prod_{l=1}^j(h^--\lambda_l)^2,\qquad j=1,2,\qquad {\cal
P}_{e^+}(h^+)\equiv-(e^+)^2,\label{peh115}\end{gather} where properties of the
polynomial ${\cal P}_{e^+}(\lambda)$ are described in Theorem~\ref{theorem3}.
Thus, the algebraic multiplicity of the energy of a bound state of
$h^-$ in the spectrum of the matrix $\bf S$ of $e^-$ is equal to~2,
i.e.\  to doubled algebraic multiplicity of this energy in the
spectrum of $h^-$ (cf.~with \eqref{peh52}).

(2) As it is known \cite{teorsol}, the Hamiltonians with
f\/inite-zone periodic potentials represent a partial case of Hamiltonians
with quasiperiodic  and, in general, complex potentials for which there
are nonzero $t$-antisymmetric symmetry operators. As well the  Hamiltonians of the type $h^-$ consi\-de\-red
above belong to the case of
Hamiltonians with potentials, which are called ``ref\/lectionless
potentials against the background of f\/inite-zone potentials'', and
for which there are nonzero $t$-antisymmetric symmetry operators
too.

One could generalize the results of the previous and present sections onto the Hamiltonians with
quasiperiodic potentials and the Hamiltonians with ref\/lectionless potentials
against the background of f\/inite-zone potentials. In particular, one can
conjecture, that in the latter case the algebraic multiplicity of the energy of
any bound state of $h$  in the spectrum of the matrix $\bf S$
of $e$ is equal to doubled algebraic multiplicity of this energy in the
spectrum of $h$ (we suppose, that the energy is not located inside or
on a border of $h$ continuous spectrum). This hypothesis is natural\footnote{The appearance
of second powers in \eqref{peh52} for cofactors corresponding to bound states is explained in \cite{cojapl08}
for a case with real-valued potentials through shrinking forbidden energy
bands.} in view of \eqref{peh}, \eqref{peh52}, \eqref{peh115} and
the fact, that ref\/lectionless potentials against the background of f\/inite-zone
potentials are limiting cases \cite{teorsol} of quasiperiodic
potentials, when some of its periods tend to inf\/inity and some allowed energy
bands shrink into points, being energies of bound states. If
this hypothesis is valid we can derive the factorization for $e$ analogous
to \eqref{os77'} and~\eqref{e114}. But the central position in the factorization will
be occupied by a nonzero non-minimizable $t$-antisymmetric symmetry operator for
the corresponding intermediate Hamiltonian with f\/inite-zone potential without
bound states. One can surmise also that the algebraic multiplicity of any border
of the continuous spectrum of $h$ (but not a border between allowed energy
bands, see Remark~\ref{remark4}), in the spectrum of the matrix $\bf S$ of $e$ is odd (cf.\
with \eqref{peh}, \eqref{peh52} and \eqref{peh109}).

\subsection*{Acknowledgments}
This work was partially supported by Grant RFBR 09-01-00145-a and Program RNP2009-1575.
The work of
A.A. was also  supported by
grants 2005SGR00564,  2009SGR, FPA2007-66665 and by the
Consolider-Ingenio 2010 Program CPAN (CSD2007-00042).

\pdfbookmark[1]{References}{ref}
\LastPageEnding

\end{document}